\documentclass[11pt,a4paper]{article}
\usepackage{jheppub,xcolor}
\pdfoutput=1
\usepackage[utf8]{inputenc}
\graphicspath{{./figs/}}

\usepackage{amssymb}
\usepackage{dcolumn}	
\usepackage{bm}			
\usepackage{bbm} 		
\usepackage{multirow}
\usepackage{slashed}
\usepackage{blkarray}
\usepackage{booktabs}
\usepackage{makecell}
\usepackage{longtable}
\usepackage{placeins}
\usepackage{graphicx}
\usepackage{amsmath}
\usepackage{float}
\usepackage[font=small,labelfont=bf]{caption}
\usepackage{etoolbox}

\definecolor{Red}{rgb}{1.,0.,0.}

\newcommand{\mcO}{\mathcal{O}}

\newcommand{\phSpa}{\ensuremath{\phantom{\frac{C^Q}{1}}}\hspace{-1.75em}}

\newcolumntype{?}[1]{!{\vrule width #1}}

\newcommand{\Ohq}{\mcO_{\varphi q}^{(1)}}
\newcommand{\Ohqt}{\mcO_{\varphi q}^{(3)}}
\newcommand{\Ohu}{\mcO_{\varphi u}}
\newcommand{\Ohd}{\mcO_{\varphi d}}

\newcommand{\chq}{c_{\varphi q}^{(1)}}
\newcommand{\chqt}{c_{\varphi q}^{(3)}}
\newcommand{\chu}{c_{\varphi u}}
\newcommand{\chd}{c_{\varphi d}}

\newcommand{\bc}{\begin{center}}
\newcommand{\ec}{\end{center}}
\newcommand{\ba}{\begin{array}}
\newcommand{\ea}{\end{array}}

\title{Improved Precision in $Vh(\rightarrow b\bar b)$\\ via Boosted Decision Trees}
\author[]{Philipp~Englert}

\emailAdd{philipp.englert@tu-dortmund.de}

\date{\today}
\abstract{
Extracting bounds on BSM operators at hadron colliders can be a highly non-trivial task. It can be useful or, depending on the complexity of the event structure, even essential to employ modern analysis techniques in order to measure New-Physics effects. A particular class of such modern methods are Machine-Learning algorithms, which are becoming more and more popular in particle physics. We attempt to gauge their potential in the study of $Vh(\rightarrow b\bar b)$ production processes, focusing on the leptonic decay channels of the vector bosons. Specifically, we employ boosted decision trees using the kinematical information of a given event to discriminate between signal and background. Based on this analysis strategy, we derive bounds on four dimension-6 SMEFT operators and subsequently compare them with the ones obtained from a conventional cut-and-count analysis. We find a mild improvement of $\mathcal{O}(\mathrm{few}\, \%)$ across the different operators.
}


\begin{document}
\begin{flushright}
\end{flushright}

\maketitle
\flushbottom

\section{Introduction}
Machine-Learning (ML) algorithms are a set of very powerful tools that really shine in situations where we have access to large amounts of data. Data is something that is usually abundant in hadron-collider experiments, so it seems natural to apply these kinds of algorithms in their study. One of their main applications in this context is the classification of events into signal and background, which can be quite hard to do well with traditional cut-and-count-based analyses or can simply be improved upon, depending on the process at hand. We would like to note, however, that there are many other important applications of ML techniques in particle physics, a detailed discussion of which is beyond the scope of this paper. For more in-depth information about the current applications of ML in high-energy physics and its advantages, we would like to refer the reader to, e.g., {refs.~\cite{Guest:2018yhq,Bourilkov:2019yoi,Schwartz:2021ftp,Karagiorgi:2021ngt}}. Furthermore, ref.~\cite{Feickert:2021ajf} attempts to curate a steadily up-to-date list of both general and more specialised reviews and recent research at the interface between ML and high-energy physics.\\
This paper accompanies our earlier study~\cite{Bishara:2022vsc0} of the $Vh(\rightarrow b\bar b)$ production processes at hadron colliders, where we derived bounds on the couplings of four dimension-6 operators within the framework of the Standard Model Effective Field Theory (SMEFT). The goal of the present study is to leverage the strengths of Machine-Learning techniques to improve the sensitivity to New-Physics effects with respect to the conventional cut-and-count approach we took in said previous analysis, and to compare the performance of the two methods.\\
Diboson production channels like this one allow EW precision measurements at hadron colliders and can be used to probe the dynamics of the Higgs boson at high energies~\cite{Farina:2016rws,deBlas:2013qqa,Franceschini:2017xkh}. Because of this, they are a viable way to test for a large class of New Physics models. For further theoretical details on $Vh$-diboson-production channels in the context of EW precision measurements, we would like to refer the reader to our previously mentioned companion paper~\cite{Bishara:2022vsc0} and our closely related earlier papers, refs.~\cite{Bishara:2020vix0,Bishara:2020pfx0}, on the corresponding diphoton channels $Wh(\rightarrow \gamma\gamma)$ and $Zh(\rightarrow \gamma\gamma)$. Previous studies on precision measurements in the $Vh$ production channel, where a boosted Higgs decaying into two $b$-quarks was considered can be found in refs.~\cite{Banerjee:2018bio,Liu:2018pkg,Banerjee:2019pks,Banerjee:2019twi,Banerjee:2021efl}. Furthermore, the ATLAS collaboration published a comprehensive study using LHC Run 2 data in refs.~\cite{ATLAS:2020fcp,ATLAS:2020jwz}. \\ 
The reason we decided to study the potential of ML techniques in the $h\rightarrow b\bar b$ channel instead of the $h\rightarrow \gamma \gamma$ channel is that, as explained in refs.~\cite{Bishara:2020vix0,Bishara:2020pfx0}, it is possible to essentially render the $h\rightarrow \gamma \gamma$ channel background-free by applying cuts on the kinematical variables. Therefore, there is not much room for improvement with an ML analysis of the diphoton channel. The situation is different for the $h\rightarrow b\bar b$ channel, though, where the backgrounds are large and more difficult to separate from the signal.\\
We decided to use BDTs for this analysis because we only have at most $\mathcal O (10)$ kinematical variables that are potentially useful for discriminating signal and background, and BDTs are relatively easy to optimise compared to, e.g., NNs. \\
We opted for performing this analysis using LHC simulations because the FCC-hh data we simulated for our previous study in ref.~\cite{Bishara:2022vsc0} is limited and as a consequence, we estimate that the statistical uncertainties are of the order of the performance difference between the BDT analysis and the conventional cut-and-count analysis. However, we want to be able to quantitatively compare the two methods with each other. We have access to better statistics for the LHC, since we performed more extensive simulations for this collider. We choose to derive the bounds assuming a luminosity of $3\,\mathrm{ab^{-1}}$, which corresponds to the HL-LHC.\\
ML algorithms have numerous advantages over more traditional algorithms. For example, they are sometimes able to recognise data patterns that are not immediately obvious using more conventional methods and hence could not be leveraged by the latter. Or they might be much simpler to set up for some types of problems where a more traditional solution would require very complex code or a lot of hand-tuning or even be outright impossible to find. However, there are arguably also some disadvantages to using an ML solution to a given problem. A common point of criticism is that it is sometimes difficult to understand in detail what the ML model is doing and why it is performing a certain way for a given data set. But there are ways of mitigating this issue by making the results more interpretable. One way to do this is the use of SHAP values, as suggested in ref.~\cite{Grojean:2022mef}.\\
The first part of section~\ref{sec:theory} is dedicated to a concise recapitulation of the general setup of the physics analysis that has been described in much more detail in ref.~\cite{Bishara:2022vsc0}. Since these prerequisites are identical to what we discussed in said companion paper, we refrain from going into much detail here. In the second part of this section, we briefly introduce the concept of the aforementioned SHAP values. In section~\ref{sec:MLsetup}, we describe the setup of the ML analysis we performed. Finally, we report our results in section~\ref{sec:MLanalysis}. On the one hand, we present the bounds derived from our ML-based analysis and compare them to the bounds derived from our previous cut-and-count analysis. On the other hand, we attempt to explain and interpret the workings of the BDTs we trained. For this, we present the aforementioned SHAP values and we visualise how the BDTs act to separate signal from background by comparing the distributions of the kinematical variables after the ML analysis with the distributions after the cut-and-count analysis and with the distributions before the application of either of the two.
\section{Theoretical background, physical and technical prerequisites}\label{sec:theory}
In this section, we briefly discuss the theoretical fundamentals related to our analysis and the physical and technical concepts we adopt from our companion paper~\cite{Bishara:2022vsc0}. In subsection~\ref{sec:prerequisites}, we concisely summarise the general setup of our analysis from a physical perspective. A more detailed discussion can be found in our companion paper. In section~\ref{sec:ShapleyValues}, we introduce the concept of SHAP values, which we will later use to explain the output of our ML model.
\subsection{General setup of the physics analysis}\label{sec:prerequisites}
\begin{figure}[t]\centering
	\includegraphics[trim={0 0 3.5cm 0},clip,scale=1]{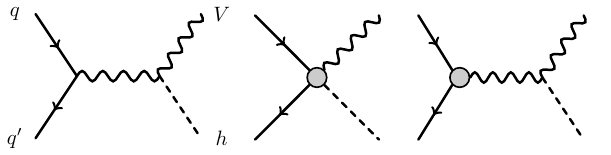}
	\caption{Tree-level Feynman diagrams contributing to the $qq'\to Vh$ production processes. The gray circles represent an insertion of one of the BSM operators in equations~\eqref{eq:Ofq1}--\eqref{eq:Ofd}.}
	\label{fig:mod_vh}
\end{figure}

Following the analysis strategy explained in ref.~\cite{Bishara:2022vsc0}, we focus on the four operators
\begin{align}
\Ohq =&\left(\overline{Q}_{L}  \gamma^{\mu} Q_{L}\right)\left(i H^{\dagger}  \overset{\leftrightarrow}{D}_{\mu} H\right)\,,\label{eq:Ofq1} \\
\Ohqt =&\left(\overline{Q}_{L} \sigma^{a} \gamma^{\mu} Q_{L}\right)\left(i H^{\dagger} \sigma^{a} \overset{\leftrightarrow}{D}_{\mu} H\right)\,, \label{eq:Ofq3} \\
\Ohu =&\left(\overline{u}_{R}\gamma^{\mu} u_R\right)\left(i H^{\dagger} \overset{\leftrightarrow}{D}_{\mu} H\right)\,, \label{eq:Ofu}\\
\Ohd =&\left(\overline{d}_{R}\gamma^{\mu} d_R\right)\left(i H^{\dagger} \overset{\leftrightarrow}{D}_{\mu} H\right)\,.\label{eq:Ofd}
\end{align}
They are responsible for the leading contributions to the energy growth of BSM corrections to the $Vh$-production processes, assuming Minimal Flavour Violation (MFV). These operators are elements of the Warsaw basis, see ref.~\cite{Grzadkowski:2010es}. In figure~\ref{fig:mod_vh}, we display representative Feynman diagrams contributing at leading order (LO) to said processes.
The $Vh(\rightarrow b\bar b)$ production processes can be divided in the following three categories depending on the number of charged leptons in the final state:
\begin{itemize}
\item \emph{Zero-lepton category}: The biggest contribution to the signal in this category comes from $pp\rightarrow Z(\rightarrow \nu \bar \nu)h(\rightarrow b \bar b)$, but the signal also receives a smaller additional contribution from $pp\rightarrow W(\rightarrow\ell\nu)h(\rightarrow b\bar b)$, whenever the charged lepton is not detected. Here, we collectively denote $\ell=e,\,\mu,\,\tau$ and $\nu=\nu_e,\,\nu_\mu,\,\nu_\tau$. The dominant contribution to the $Wh$ signal in this category is the one where $W\rightarrow\tau\nu_\tau$. The backgrounds we consider in this analysis are $t\bar t$-, $Wb\bar b$- and $Zb\bar b$ production with the same $W$- and $Z$ decay modes as in the signal.
    \item \emph{One-lepton category}: The signal process in this category is ${pp\rightarrow W(\rightarrow\ell\nu)h(\rightarrow b\bar b)}$, where we denote $\nu=\nu_e,\,\nu_\mu$ and the charged lepton $\ell=e,\,\mu$ is detected. The backgrounds we include in our analysis are $Wb\bar b$ and $t\bar t$.
    \item \emph{Two-lepton category}: In this category, the signal is the $pp\rightarrow Z(\rightarrow \ell^+ \ell^-)h(\rightarrow b \bar b)$ process, where $\ell=e,\, \mu$ and both leptons are detected and the only background we consider is the $Zb\bar b$ process.
\end{itemize}
As discussed in refs.~\cite{Bishara:2020pfx0, Bishara:2022vsc0, Bishara:2019iwh}, the phase-space region that yields the highest sensitivity to the BSM operators is the high-energy tail. The Higgs boson is usually boosted in this region, which entails characteristic kinematical properties that can be tested for in order to distinguish signal- from background events. A possible way to test for such a boosted Higgs is the mass-drop-tagging technique~\cite{Butterworth:2008iy}. In order to make use of the available events in the most efficient way, we divide them into two classes: events that contain such a boosted Higgs candidate, which we call `boosted events', and `resolved events', which contain a pair of resolved $b$-jets instead. We explain this classification procedure in detail in ref.~\cite{Bishara:2022vsc0}. Since in our analysis, we combine these two classes of events, the tagging procedure we employ is a scale invariant one~\cite{Gouzevitch:2013qca}. The concrete tagging strategy we follow in our studies is heavily inspired by the one implemented in ref.~\cite{Bishara:2016kjn}. As a result, we can divide all the events into six categories depending on whether the final state contains 0, 1 or 2 charged leptons and depending on whether the event is boosted or resolved.\\
In our companion study, we optimised cuts on the kinematical variables of each of these six categories separately in order to distinguish signal- from background events. In the study described here, we simply use BDTs for this discrimination instead.\\
The parton-level events were generated using \texttt{MadGraph5\_aMC@NLO} v.2.7.3~\cite{Alwall:2014hca} together with the \texttt{NNPDF23} parton-distribution functions~\cite{Ball:2013hta}. The subsequent parton showers and the Higgs decay were modelled via \texttt{Pythia8.24}~\cite{Sjostrand:2014zea}. The SMEFT operators were defined and integrated into our analysis using the \texttt{SMEFTatNLO} \texttt{UFO} model~\cite{Degrande:2020evl,Degrande:2011ua}. More technical details about the simulations are explained in ref.~\cite{Bishara:2022vsc0}.
\subsection{SHAP values}\label{sec:ShapleyValues}
In order to make the predictions of our BDTs more transparent, we make use of the so-called SHAP (SHapley Additive exPlanations) values~\cite{NIPS2017_7062}, which are based on the game-theoretic concept of Shapley values~\cite{shapley1951notes}.\\
Both Shapley values and SHAP values are feature-importance measures, i.e. their goal is to quantify the influence of a certain input feature on the outcome of the prediction for a given model. \\
We can translate the game-theoretic philosophy behind Shapley values into the context of ML in the following way: the importance of an input feature can be defined as the effect of including that feature as an input on the prediction of the model. There are several ways to technically implement a quantitative measure for this effect and perhaps the most intuitive one is the approach of Shapley regression values~\cite{Lipovetsky2001}. The idea behind this is to train a model with the feature present in the input data and another model with the feature absent and compare the predictions of the two models. \\There is one caveat to this idea, though, which is that the effect of including versus withholding the feature in the input data depends on which other features are included in the input data. If, for example, there is another input feature that is highly correlated with the feature whose importance we are trying to determine, then the importance is shared between those two input features. Since this should be reflected by a quantitative importance measure, Shapley values are computed in the following way.\\
Let $i$ be the input feature whose importance we want to measure and $F$ be the set of all the input features we are using. We call $S$ a subset of the set of all the input features excluding the one whose importance we want to measure, i.e. $S\subseteq F\backslash \{i\}$. Now we train a model $f_{S\cup \{i\}}(x_{S\cup\{i\}})$ and a model $f_S(x_S)$ for each possible subset $S$, including the empty set $\O$. Here, $x_{S\cup\{i\}}$ and $x_S$ are the input feature values of a specific sample $x$, restricted to the subset $S$ of the features with and without the feature $i$, respectively. The Shapley regression value $\phi_i(x)$ for the sample $x$ of the input feature $i$ is defined as
\begin{align}
    \phi_i(x) = \sum\limits_{S\subseteq F\backslash \{i\}} \frac{|S|!(|F|-|S|-1)!}{|F|!}[f_{S\cup \{i\}}(x_{S\cup\{i\}}) - f_S(x_S)]\, .
\end{align}
The normalisation factor that weights the contributions is closely related to the inverse number of subsets of input features of size $|S|$ that can be chosen from the set of all the input features excluding $i$, which is of size $|F|-1$. For example, the contribution from the model where $S$ is the full set, $S=F\backslash\{i\}$, is much more important than the contribution from one of the models where half of the input features are in $S$. This is to account for the fact that if we did not apply the weights in that way, the effect of, e.g., a strong correlation of $i$ with another input variable on the Shapley value would be diluted if there was a large number of other input variables, even if they did not contribute much to the prediction at all.\\
To give a more applied example, imagine the problem of discriminating between the ${Zh(\rightarrow b\bar b)}$ and ${Wh(\rightarrow b\bar b)}$ signal events of the resolved 0-lepton category and the background consisting of $Zb\bar b$-, $Wb\bar b$- and $t\bar t$ events. Let the input variables be the mass $m_{bb}$ of the $b\bar b$ pair, the missing transverse momentum $E_T^\mathrm{miss}$ and the angular distance between the two $b$-jets $\Delta R_{bb}$. Of course, this is not an optimal selection of input variables, but it serves us well to illustrate the point. \\
To assess the importance of $m_{bb}$ for the prediction of a specific event, we train a model with only $\Delta R_{bb}$ as an input variable and compare its prediction for this event with the one of the model with $\Delta R_{bb}$ and $m_{bb}$ as an input variable. Multiplying this difference with the according combinatorial weight gives us the contribution of $S=\{\Delta R_{bb}\}$ to the Shapley value. Now, we do the same with $S=\{E_T^\mathrm{miss}\}$, $S=\{E_T^\mathrm{miss},\, \Delta R_{bb}\}$ and $S=\{\O\}$, weight them accordingly and add up the contributions. The result is the Shapley value of $m_{bb}$ for the given event.\footnote{Note that for $S=\{\O\}$, we need to define the output of the model with no input variable $f_{\O}(x_{\O})$. The only information available to train this model is the distribution of the true labels of the training set, so we define $f_{\O}(x_{\O})$ to be the expectation value of the true label over the training set.} Since, the discriminating power of $m_{bb}$ in this process is large, an ML model that performs well would likely yield a large Shapley value for $m_{bb}$. The same can be done to compute the Shapley values of $\Delta R_{bb}$ and $E_T^\mathrm{miss}$.\\
This procedure, however, is very computationally expensive because a new model needs to be trained for every possible subset of the input variables. In fact, this problem is even NP-hard.\\
A different approach, the so-called SHAP (SHapley Additive exPlanations) value~\cite{NIPS2017_7062}, unifies the ideas of several methods that approximate the Shapley regression values with the ideas of several other feature-importance measures and circumvents this issue of having to train a model for every possible subset of the input variables. Instead, one measures the effect of removing a variable from the model by averaging the predictions of the model when drawing values for the variable to be removed from random samples of the data set. I.e., one uses the expectation function of the model conditional on the variables that have not been removed. \\
Since even SHAP values are still computationally very expensive, one can apply different approximations, e.g. one can also sample from the feature subsets instead of taking into account all of them, which is known as Shapley sampling values~\cite{Strumbelj2013}. Alternatively, if one focuses on specific types of models, e.g. decision trees, one can make use of their properties to use algorithms that are able to exactly compute the classic Shapley regression values in polynomial time. One of these algorithms is the \textsc{TreeExplainer}~\cite{lundberg2019explainable,2018arXiv180203888L}, which is used in this work to measure the feature importance in BDTs.
\section{Setup of the analysis}\label{sec:MLsetup}
In this section, we present the methodology of the ML analysis replacing the conventional cut-and-count analysis described in ref.~\cite{Bishara:2022vsc0}. \\
We use one BDT per bin, according to the definitions in table~\ref{tab:bin_boundarieshbb}, for each of the 0-, 1- and 2-lepton categories and separately for both boosted and resolved events. Furthermore, we treat the extracted observables to be independent in the context of the computation of the $\chi^2$ function. We decided to perform this bin-by-bin analysis for the following reason. The sensitivity to the New-Physics operators varies significantly between the different bins, and this approach avoids the potential problem that the performance of the ML model could be optimised for a relatively unimportant bin at the cost of the performance in a bin with larger sensitivity.
\begin{table}[t]
	\centering{
		\renewcommand{\arraystretch}{1.25}
		\scalebox{0.85}{
		\begin{tabular}{  c | c | c | c  }
		\toprule
		\multicolumn{2}{c|}{Categories} & Variable & bin boundaries  \\
		\midrule
		 & boosted & \multirow{2}*{$p_{T,\mathrm{min}}\, \mathrm{[GeV]}$ } & $\{0, 300, 350,\infty\}$  \\
		\multirow{-1}{*}[1.9ex]{0-lepton} & resolved & &$\{0, 160, 200, 250,\infty\}$  \\
		\midrule
		& boosted & \multirow{2}*{$p_{T}^h \, \mathrm{[GeV]}$} & $\{0, 175, 250, 300, \infty\}$  \\
		\multirow{-1}{*}[1.9ex]{1-lepton} & resolved & &$\{0, 175, 250,\infty\}$  \\
		\midrule
		& boosted & \multirow{2}*{$p_{T,\mathrm{min}} \, \mathrm{[GeV]}$ } &$\{250,\infty\}$  \\
		\multirow{-1}{*}[1.9ex]{2-lepton} & resolved & & $\{175, 200,\infty\}$  \\
		\bottomrule
		\end{tabular}}}
		\caption{Bin definitions used in the different categories of the analyses for the HL-LHC, adopted from ref.~\cite{Bishara:2022vsc0}.}
		\label{tab:bin_boundarieshbb}
\end{table}\\
We preselect the events that are used for each of the individual BDT analyses according to the number of charged leptons that pass the acceptance cuts and whether the event contains a boosted Higgs candidate or two resolved $b$-jets. Furthermore, we do not accept events where the Higgs candidate does not have two $b$-tags and for all but the boosted 2-lepton category we veto events with untagged jets inside the acceptance region. The identical preselection cuts are implemented in the conventional cut-and-count analysis and only events that pass these criteria are used for the respective ML analyses.\\
As a loss function, we used the binary cross-entropy, which can be defined via~\cite{Mehta:2018dln}
\begin{align}
    H_p(x) = -\frac{1}{N} \sum_{i=1}^{N} w_i \bigg[ y_i \cdot \log\big(p\left(x_i\right)\big)+(1-y_i)\cdot\log\big(1-p\left(x_i\right)\big)\bigg]\, ,
\end{align}
where $N$ is the number of samples being used for the computation of the loss function, $y_i\in\{0,1\}$ is the true class label (signal or background) of the $i$-th sample and $x_i$ are the features of the $i$-th sample, which, in the case of this study, are the kinematical variables the predictions are based on. The prediction the model in question makes for the $i$-th sample is denoted as $p(x_i)$. Each sample can be assigned an individual weight $w_i$ that quantifies its importance for the performance of the model. In order to account for the differential cross-section $\sigma_i$ associated to each event, we define the weights to be
\begin{align}
    w_i = \frac{\tilde{w}_i \cdot \sigma_i}{\langle \tilde{w}_j \cdot \sigma_j \rangle\Big|_{j=1}^{N}}\, ,
\end{align}
where $\langle . \rangle\Big|_{i=1}^{N}$ denotes the mean value over the $N$ samples and
\begin{align}
    \tilde{w}_i = \begin{cases}
    w_\mathrm{sig} \, &\text{ for } y_i = 1\\
    1 \, &\text{ for } y_i = 0\,
    \end{cases}
\end{align}
assigns a relative weight $w_\mathrm{sig}$ to signal events compared to background events in order to be able to maximise
\begin{align}
    \frac{s}{\sqrt{b}} :=\frac{\mathrm{signal}}{\sqrt{\mathrm{background}}}\, ,
\end{align}
instead of just $s/b$. For this, the signal weight $w_\mathrm{sig}$ is to be tuned during the stage of hyperparameter optimisation.\\
The hyperparameters \textit{signal weight}, \textit{maximum depth} of the individual decision trees and \textit{learning rate} were optimised via $5$-fold cross-validation (CV) for each individual BDT used in the analysis. For each bin, $n$-lepton- and boosted/resolved-category, we performed an exhaustive scan of the grid defined by the following hyperparameter values:
\begin{align}\label{eq:hyperparametergrid}
    \begin{split}
        \text{maximum depth}&=\{5, \,10,\, 15,\, 20\}\\
        \text{learning rate}&=\{10^{-3},\, 10^{-4}\}\\
        \text{signal weight}&=\{0.5, \,1, \,5, \,10, \,30,\, 50, \,100, \,500, \,1000\}\, .
    \end{split}
\end{align}
After each of these grid searches, we picked the configuration with the largest value of $s/\sqrt{b}$ averaged over the five iterations of CV under the side condition that the variance be moderate. \\
During the grid search, the maximum number of trees for each BDT was set to $10^5$. An optimisation of this parameter was not necessary because we used early-stopping as a regularisation method, which automatically truncates the number of BDTs at the optimal value. For the BDT of the boosted 2-lepton category and the last bin of the resolved 2-lepton category, however, the early stopping condition was not met within those $10^5$ training steps, so for those two exceptions, we decided to set the maximum number of trees to $10^6$ during the final training with the optimised hyperparameters. However, due to limited computational resources, we could not afford to do a full grid search with this higher maximum number of trees during the hyperparameter-optimisation stage.\\
We randomly split the total amount of data into training set (60\%), test set (30\%) and validation set (10\%). This random assignment of events into training-, test- and validation set was performed using stratified sampling, i.e., the random sampling was subject to the side condition that each physical process (e.g. $Wh$, $Wb\bar b$ and $t\bar t$ for the 1-lepton category) should be represented in equal proportions in each of the of the three data sets.
\section{Results}\label{sec:MLanalysis}
In this section, we present the results we obtained from our final analysis using the BDTs with the optimised sets of hyperparameters displayed in table~\ref{tab:hyperparams}. 
\begin{table}[b!]
\begin{centering}
\setlength{\extrarowheight}{0mm}%
\scalebox{0.95}{
\begin{tabular}{c|c?{1.5pt}c|c|c}
\toprule[1.5pt]
\rule[-.5em]{0pt}{.5em}
Category & $p_T$ bin [GeV] & max. depth  & learning rate & signal weight\tabularnewline
\midrule[1.5pt]
\multirow{3}{*}{\rule{0pt}{41pt}  $\begin{aligned}&\text{0 - lepton}\\&\text{resolved}\end{aligned}$} & $[0,\,160]$ & \rule{0pt}{1.3em} 15 & $10^{-3}$ & 100 \tabularnewline[0.3em]
\cline{2-5} 
 & $[160,\,200]$ & \rule{0pt}{1.3em} 5 & $10^{-3}$ & 30  \tabularnewline[0.3em]
\cline{2-5} 
 & $[200,\,250]$ & \rule{0pt}{1.3em} 5 & $10^{-3}$ & 50  \tabularnewline[0.3em]
 \cline{2-5} 
 & $[250,\,\infty]$ & \rule{0pt}{1.3em} 10 & $10^{-2}$  & 500  \tabularnewline[0.3em]
\hline
\multirow{3}{*}{\rule{0pt}{30pt} $\begin{aligned}&\text{0 - lepton}\\&\text{boosted}\end{aligned}$} &  $[0,\,300]$ &  \rule{0pt}{1.3em} 15 & $10^{-3}$ & 500\tabularnewline[0.3em]
\cline{2-5} 
 & $[300,\,350]$ & \rule{0pt}{1.3em} 5 & $10^{-2}$ & 10  \tabularnewline[0.3em]
\cline{2-5} 
 &$[350,\,\infty]$ & \rule{0pt}{1.3em} 5 & $10^{-2}$ & 10 \tabularnewline[0.3em]
\hline
\multirow{3}{*}{\rule{0pt}{30pt} $\begin{aligned}&\text{1 - lepton}\\&\text{resolved}\end{aligned}$} & $[0,\,175]$ &  \rule{0pt}{1.3em} 5 & $10^{-3}$ & 100 \tabularnewline[0.3em]
\cline{2-5} 
 & $[175,\,250]$ & \rule{0pt}{1.3em} 5 & $10^{-2}$ & 30\tabularnewline[0.3em]
\cline{2-5} 
 & $[250,\,\infty]$ & \rule{0pt}{1.3em} 5 & $10^{-3}$ & 1000  \tabularnewline[0.3em]
\hline
\multirow{3}{*}{\rule{0pt}{41pt}  $\begin{aligned}&\text{1 - lepton}\\&\text{boosted}\end{aligned}$} & $[0,\,175]$ &  \rule{0pt}{1.3em} 5 & $10^{-3}$ & 1000 \tabularnewline[0.3em]
\cline{2-5} 
 & $[175,\,250]$ & \rule{0pt}{1.3em} 10 & $10^{-3}$ & 100 \tabularnewline[0.3em]
\cline{2-5} 
 & $[250,\,300]$ & \rule{0pt}{1.3em} 10 & $10^{-2}$ & 100\tabularnewline[0.3em]
 \cline{2-5} 
 & $[300,\,\infty]$ & \rule{0pt}{1.3em} 15 & $10^{-3}$ & 50 \tabularnewline[0.3em]
\hline
\multirow{2}{*}{\rule{0pt}{20pt}  $\begin{aligned}&\text{2 - lepton}\\&\text{resolved}\end{aligned}$} & $[175,\,250]$ &  \rule{0pt}{1.3em} 10 & $10^{-3}$ & 500  \tabularnewline[0.3em]
\cline{2-5} 
 & $[250,\,\infty]$ & \rule{0pt}{1.3em} 5 & $10^{-3}$ & 100 \tabularnewline[0.3em]
\hline
\rule{0pt}{21pt}$\begin{aligned}&\text{2 - lepton}\\&\text{boosted}\end{aligned}$ & $[250,\,\infty]$ &  \rule{0pt}{1.3em} 10& $10^{-3}$ & 30\tabularnewline[0.3em]
 
\bottomrule[1.5pt]
\end{tabular}
}
\par\end{centering}
\caption{Optimal sets of hyperparameters for the BDTs given bin-by-bin for each category, determined by performing a grid search defined by the hyperparameter values given in equation~\eqref{eq:hyperparametergrid}. The sets of hyperparameters we consider to be optimal are the ones that yield the largest average value of $s/\sqrt{b}$ over five CV folds under the side condition of not having an excessively large variance.}
\label{tab:hyperparams}
\end{table}
On the one hand, we compute the bounds derived from the BDT-based analysis of the $Vh(\rightarrow b \bar b)$ process, which we will do in section~\ref{sec:boundsML}.\\
On the other hand we would like to shine some light on how the BDTs use the kinematical variables of the events to discriminate between signal and background. For this purpose, we make use of the previously introduced SHAP values and to complement this, we show some of the kinematical distributions of both signal and background events before and after both the BDT-based analysis and the cut-and-count analysis. These aspects of explainability and interpretability of the ML model are explored in section~\ref{sec:explainabilityML}. \\
Lastly, we draw conclusions from these results and attempt to use them to highlight the advantages and disadvantages of an ML-based approach over a conventional cut-and-count approach in section~\ref{sec:conclusionML}.
\subsection{Presentation of the bounds}\label{sec:boundsML}
Like in the conventional cut-and-count analysis discussed in our previous paper~\cite{Bishara:2022vsc0}, we generate histograms using the events that have been accepted by the BDTs according to the bin definitions given in table~\ref{tab:bin_boundarieshbb}. For the signal, we fit one-dimensional quadratic functions of the Wilson coefficients $\chqt$, $\chq$, $\chu$ and $\chd$ to this data in a bin-by-bin fashion. From these bin-by-bin fits and the background histograms, we construct one-dimensional $\chi^2$ functions of the Wilson coefficients in order to derive the $95\%$ CL bounds at $\chi^2=3.84$. The results of these fits and the background histograms are given in tables~\ref{tab:App_sigma_full_Zh_neut_HL_LHC_res_BDT}~-~\ref{tab:App_sigma_full_Zh_lep_HL_LHC_boos_BDT} in appendix~\ref{sec:app_tables_ML}.\\
In table~\ref{tab:bounds_summary_HL_LHC_ML}, we present the bounds we derived for the three benchmarks of $1\%$, $5\%$ and $10\%$ systematic uncertainty. All these bounds represent an improvement over the bounds obtained via the conventional cut-and-count analysis discussed in our companion paper. The relative improvements of both the lower and upper bounds are also given in table~\ref{tab:bounds_summary_HL_LHC_ML} and they are mostly of the order $\mathcal{O}(\mathrm{few}\,\%)$.\\
We find the largest relative improvements in the lower bound on the $\chq$ coefficient and the weakest improvements are found for the lower bound on the $\chd$ coefficient. For most of the coefficients, the relative improvement grows with increasing systematic uncertainties. The exception to this is the $\chqt$ operator, where we find the largest improvement for the $1\%$ systematic-uncertainty benchmark.

\begin{table}[t!]
\begin{centering}
\begin{small}
\begin{tabular}{c|c|c|c}
\toprule
\multirow{2}{*}{\rule{0pt}{1.4em}Coefficient} &\multirow{2}{*}{\rule{0pt}{1.4em}Bounds$\,$[TeV$^{-2}$]} & \multicolumn{2}{c}{\rule{0pt}{1.1em}Relative improvement} \tabularnewline[0.3em] \cline{3-4}
 & & \rule{0pt}{1.3em} Lower bound &  Upper bound \tabularnewline
\midrule
$c_{\varphi q}^{(3)}\,$ &
\begin{tabular}{ll}
\rule{0pt}{1.25em}$[-1.1,\,1.0]\times10^{-2}$ & $1\%$ syst.\\
\rule{0pt}{1.25em}$[-1.7,\,1.5]\times10^{-2}$ & $5\%$ syst.\\
\rule[-.65em]{0pt}{1.9em}$[-2.9,\,2.2]\times10^{-2}$ & $10\%$ syst.
\end{tabular}
&
\begin{tabular}{ll}
&\rule{0pt}{1.25em}$5\%$   \\
&\rule{0pt}{1.25em}$1.5\%$    \\
&\rule[-.65em]{0pt}{1.9em}$1.5\%$    
\end{tabular}
& 
\begin{tabular}{ll}
&\rule{0pt}{1.25em}5\%   \\
&\rule{0pt}{1.25em}$2\%$    \\
&\rule[-.65em]{0pt}{1.9em}$3\%$    
\end{tabular}
\tabularnewline

\hline
$c_{\varphi q}^{(1)}\,$ &
\begin{tabular}{ll}
\rule{0pt}{1.25em}$[-4.2,\,6.2]\times10^{-2}$ & $1\%$ syst.\\
\rule{0pt}{1.25em}$[-5.0,\,7.0]\times10^{-2}$ & $5\%$ syst.\\
\rule[-.65em]{0pt}{1.9em}$[-6.4,\,8.3]\times10^{-2}$ & $10\%$ syst.
\end{tabular}
&
\begin{tabular}{ll}
&\rule{0pt}{1.25em}$7\%$  \\
&\rule{0pt}{1.25em}$7\%$  \\
&\rule[-.65em]{0pt}{1.9em}$7\%$  
\end{tabular}
&
\begin{tabular}{ll}
&\rule{0pt}{1.25em}$2\%$  \\
&\rule{0pt}{1.25em}$3\%$  \\
&\rule[-.65em]{0pt}{1.9em}$4\%$  
\end{tabular}
\tabularnewline

\hline
$c_{\varphi u}\,$ &
\begin{tabular}{ll}
\rule{0pt}{1.25em}$[-10.7,\,3.8]\times10^{-2}$ & $1\%$ syst.\\
\rule{0pt}{1.25em}$[-11.9,\,4.8]\times10^{-2}$ & $5\%$ syst.\\
\rule[-.65em]{0pt}{1.9em}$[-13.6,\,6.5]\times10^{-2}$ & $10\%$ syst.
\end{tabular}
&
\begin{tabular}{ll}
&\rule{0pt}{1.25em}$4\%$ \\
&\rule{0pt}{1.25em}$5\%$  \\
&\rule[-.65em]{0pt}{1.9em}$5\%$  \\
\end{tabular}
&
\begin{tabular}{ll}
&\rule{0pt}{1.25em}$2\%$ \\
&\rule{0pt}{1.25em}$3\%$  \\
&\rule[-.65em]{0pt}{1.9em}$4\%$  \\
\end{tabular}
\tabularnewline

\hline
$c_{\varphi d}\,$ &
\begin{tabular}{ll}
\rule{0pt}{1.25em}$[-6.5,\,10.2]\times10^{-2}$ & $1\%$ syst.\\
\rule{0pt}{1.25em}$[-7.9,\,11.5]\times10^{-2}$ & $5\%$ syst.\\
\rule[-.65em]{0pt}{1.9em}$[-10.1,\,13.7]\times10^{-2}$ & $10\%$ syst.
\end{tabular}
&
\begin{tabular}{ll}
&\rule{0pt}{1.25em}$0.9\%$    \\
&\rule{0pt}{1.25em}$1.6\%$   \\
&\rule[-.65em]{0pt}{1.9em}$2.6\%$   \\
\end{tabular} 
&
\begin{tabular}{ll}
&\rule{0pt}{1.25em}$4\%$    \\
&\rule{0pt}{1.25em}$4\%$   \\
&\rule[-.65em]{0pt}{1.9em}$5\%$   \\
\end{tabular}\\
\bottomrule

\end{tabular}
\end{small}
\par\end{centering}
\caption[caption]{ Bounds at $95\%$ CL ($\chi^2=3.84$)~on the coefficients of the $\Ohqt$, $\Ohq$, $\Ohu$ and $\Ohd$ operators for 14 TeV HL-LHC with integrated luminosity of $3\,\mathrm{ab}^{-1}$ using the BDT-based analysis. In the third and fourth column, we present the relative improvements of these bounds with respect to the corresponding results derived from the cut-and-count approach.
}
\label{tab:bounds_summary_HL_LHC_ML}
\end{table}
\subsection{Explaining and interpreting the results}\label{sec:explainabilityML}
\begin{figure}[t!]
    \centering
    \includegraphics[width=\linewidth]{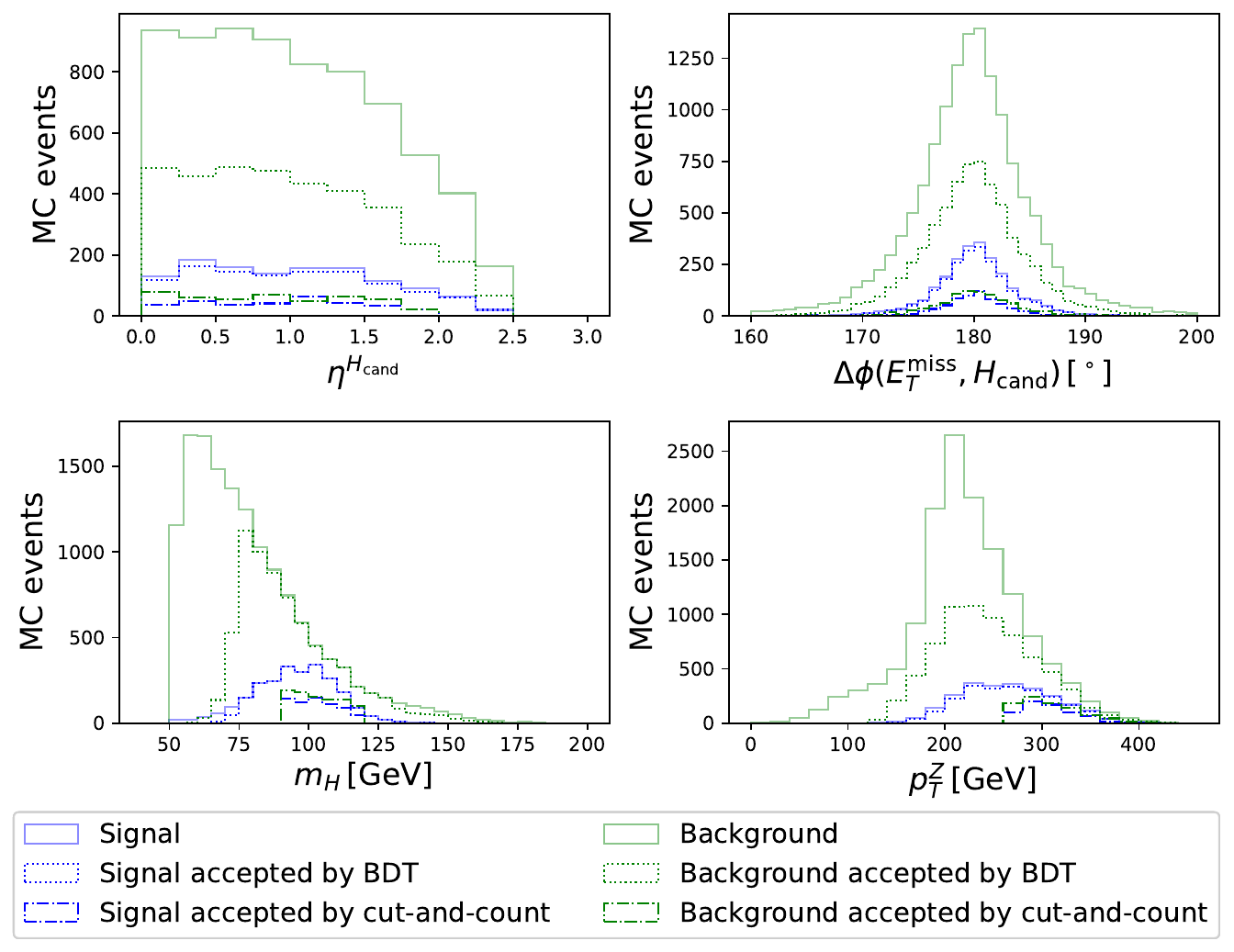}
    \caption{Kinematical distributions of both signal- and background events in the ${p_{T,\mathrm{min}}\in[0,\,300]\,}$GeV bin of the boosted 0-lepton category. The numbers of events are given as unweighted Monte-Carlo events. The signal distributions are displayed in blue, while the background distributions are displayed in green. The solid lines represent the distributions after the application of preselection cuts like the jet veto and the charged lepton veto, but before applying any other cuts or BDT analysis. The dotted lines display the distributions after the BDTs have been applied to reject events. The dash-dotted lines are associated to the distributions after the conventional cuts used in the analysis discussed in ref.~\cite{Bishara:2022vsc0}. In the top-left panel, we present the distribution of $\eta^{H_\mathrm{cand}}$, in the top-right panel $\Delta\phi(E_T^\mathrm{miss},H_\mathrm{cand})$, in the bottom-left panel $m_H$ and in the bottom-right panel $p_T^Z$.}
    \label{fig:BDTdistros_010}
\end{figure}

\begin{figure}[t!]
    \centering
    \includegraphics[width=\linewidth]{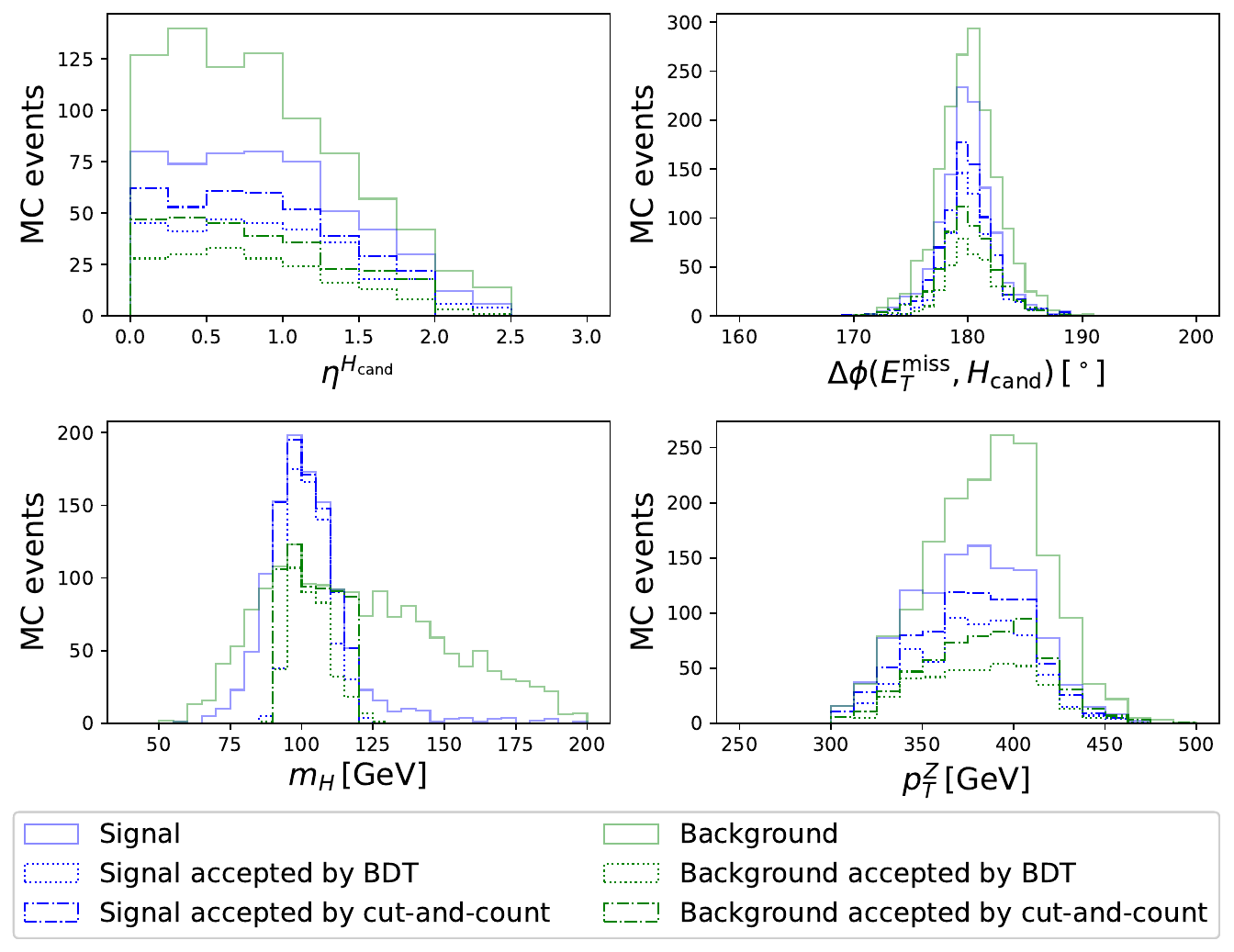}
    \caption{Kinematical distributions of both signal- and background events in the ${p_{T,\mathrm{min}}\in[300,\,350]\,}$GeV bin of the boosted 0-lepton category. The numbers of events are given as unweighted Monte-Carlo events. The signal distributions are displayed in blue, while the background distributions are displayed in green. The solid lines represent the distributions after the application of preselection cuts like the jet veto and the charged lepton veto, but before applying any other cuts or BDT analysis. The dotted lines display the distributions after the BDTs have been applied to reject events. The dash-dotted lines are associated to the distributions after the conventional cuts used in the analysis discussed in ref.~\cite{Bishara:2022vsc0}. In the top-left panel, we present the distribution of $\eta^{H_\mathrm{cand}}$, in the top-right panel $\Delta\phi(E_T^\mathrm{miss},H_\mathrm{cand})$, in the bottom-left panel $m_H$ and in the bottom-right panel $p_T^Z$.}
    \label{fig:BDTdistros_011}
\end{figure}

\begin{figure}[t!]
    \centering
    \includegraphics[width=\linewidth]{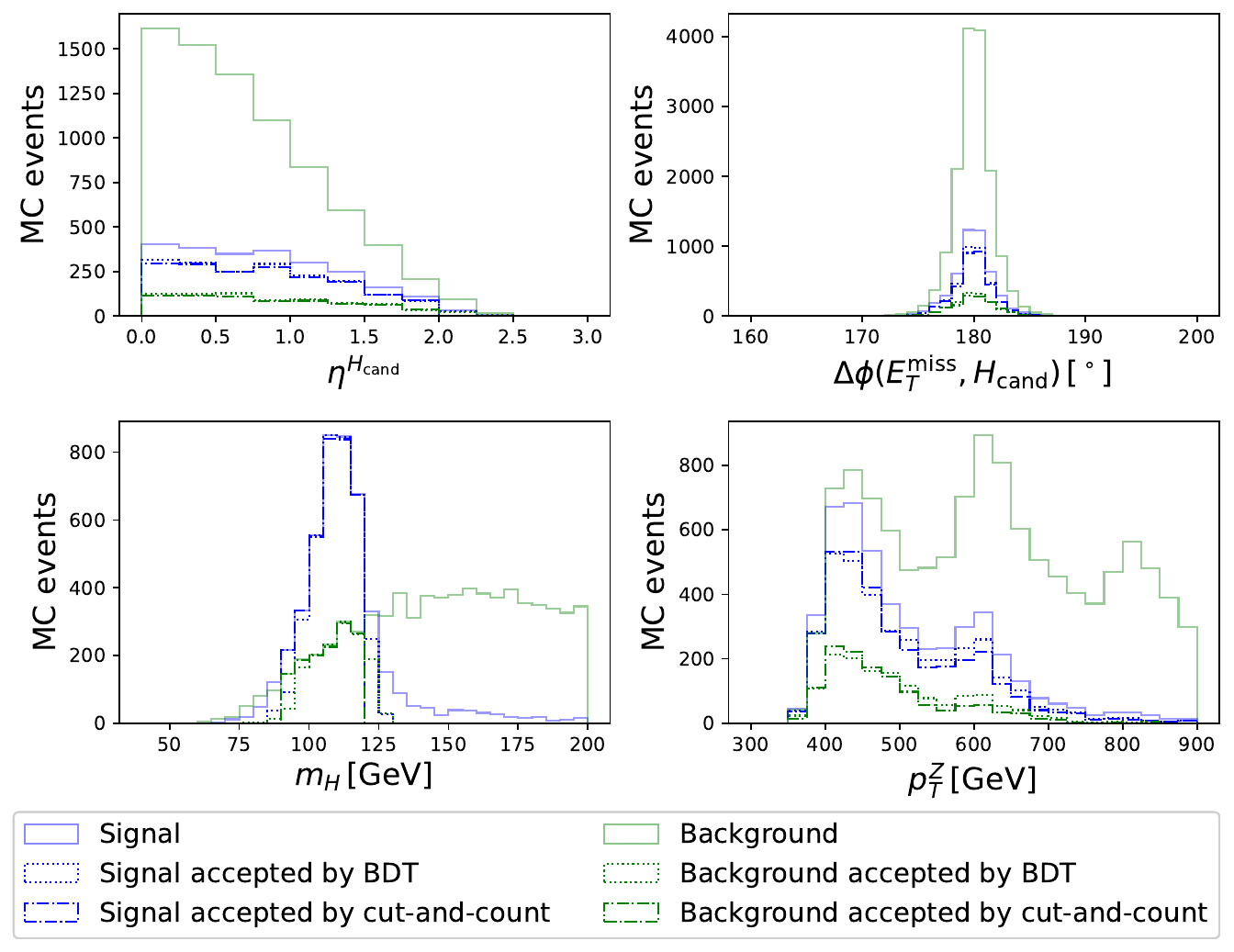}
    \caption{Kinematical distributions of both signal- and background events in the ${p_{T,\mathrm{min}}\in[350,\,\infty]\,}$GeV bin of the boosted 0-lepton category. The numbers of events are given as unweighted Monte-Carlo events. The signal distributions are displayed in blue, while the background distributions are displayed in green. The solid lines represent the distributions after the application of preselection cuts like the jet veto and the charged lepton veto, but before applying any other cuts or BDT analysis. The dotted lines display the distributions after the BDTs have been applied to reject events. The dash-dotted lines are associated to the distributions after the conventional cuts in the analysis discussed in ref.~\cite{Bishara:2022vsc0}. In the top-left panel, we present the distribution of $\eta^{H_\mathrm{cand}}$, in the top-right panel $\Delta\phi(E_T^\mathrm{miss},H_\mathrm{cand})$, in the bottom-left panel $m_H$ and in the bottom-right panel $p_T^Z$.}
    \label{fig:BDTdistros_012}
\end{figure}

\begin{figure}[t!]
	\centering
	\includegraphics[width=0.485\linewidth]{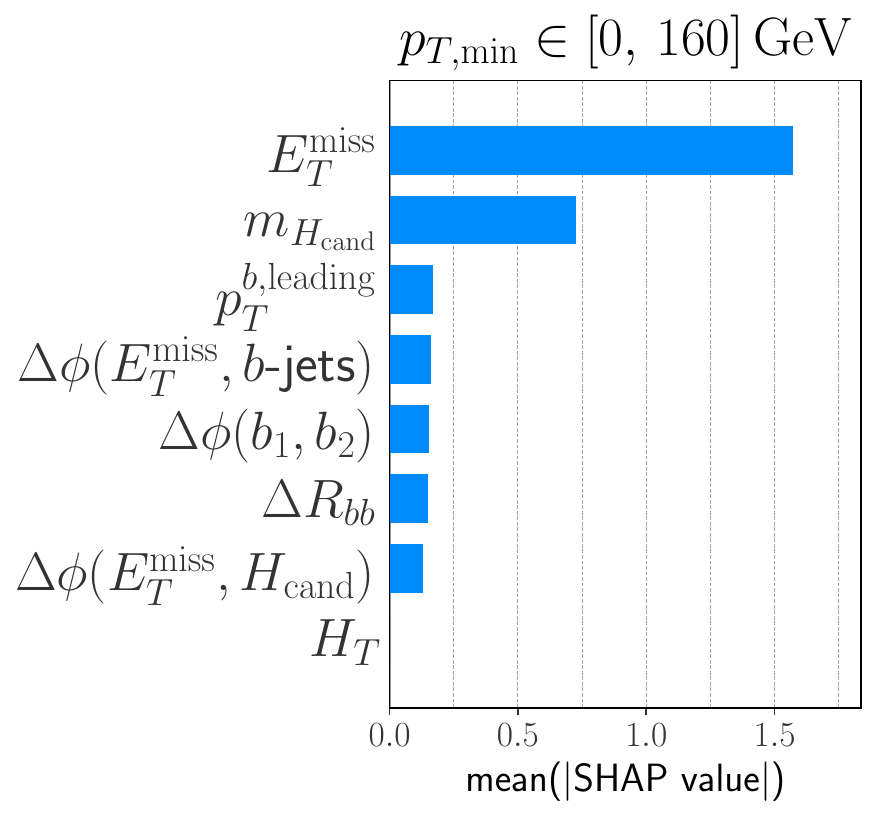} \hfill
	\includegraphics[width=0.485\linewidth]{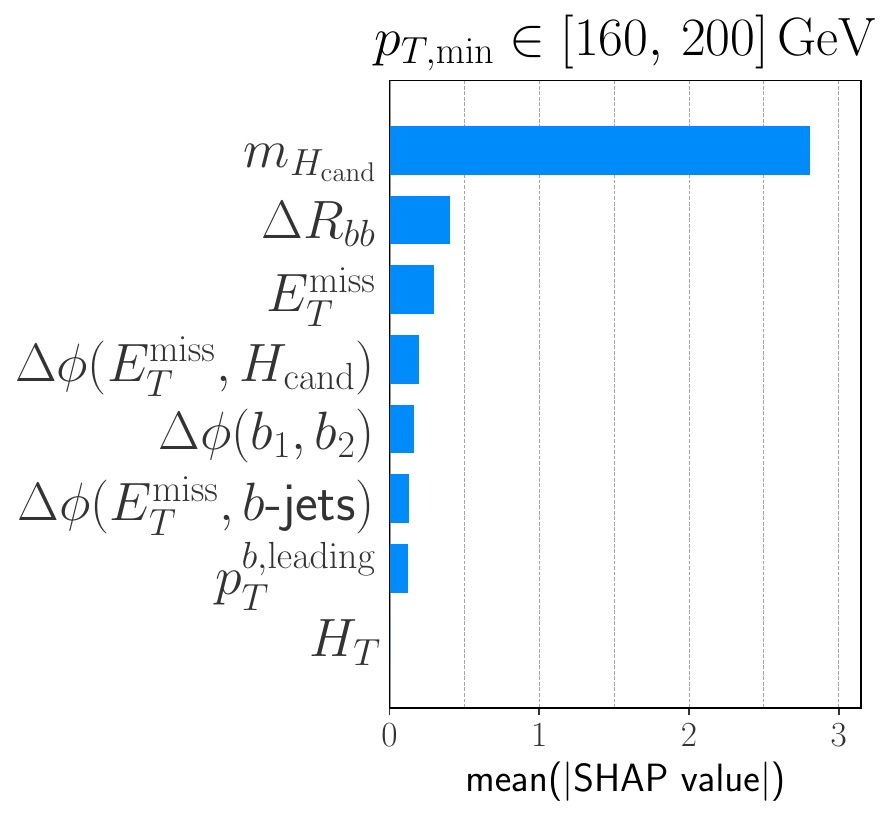} \\
	\vspace{0.3cm}
    \includegraphics[width=0.485\linewidth]{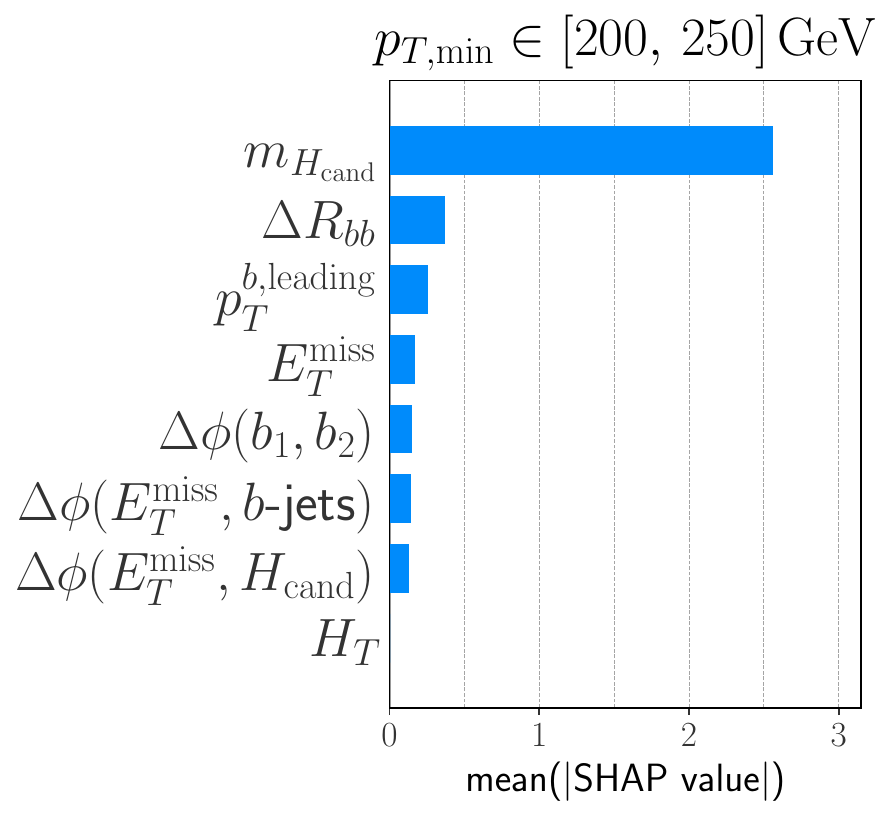}\hfill
	\includegraphics[width=0.485\linewidth]{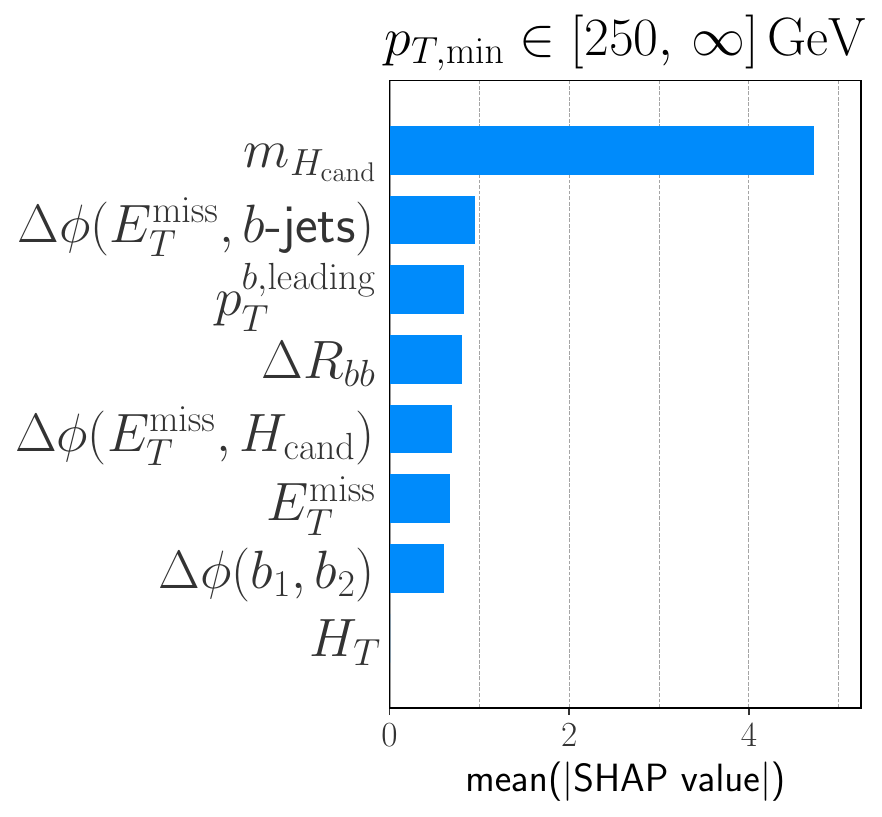}
 	\caption{Averages of the absolute SHAP values of the different input features used by the BDTs in the resolved 0-lepton category. The top-left panel corresponds to the $[0,\,160]\,\mathrm{GeV}$ bin, the top-right panel to the $[160,\,200]\,\mathrm{GeV}$ bin, the bottom-left panel to the $[200,\,250]\,\mathrm{GeV}$ bin and the bottom-right panel to the $[250,\,\infty]\,\mathrm{GeV}$ bin.}
	\label{fig:shap_00} 
\end{figure}

\begin{figure}[tb!]
	\centering
	\includegraphics[width=0.485\linewidth]{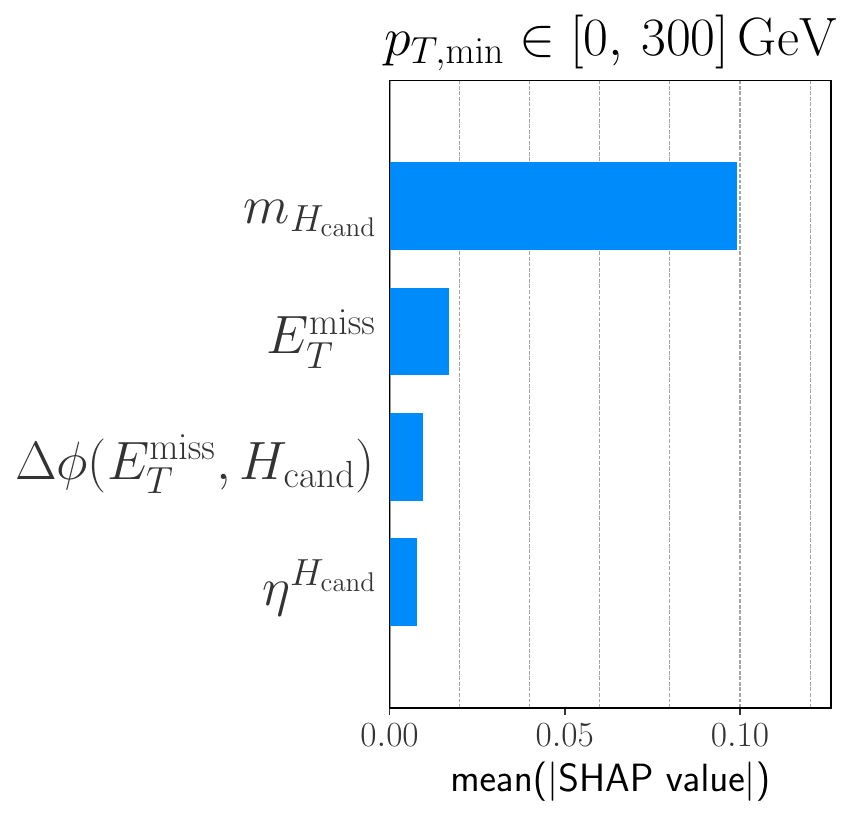} \hfill
	\includegraphics[width=0.485\linewidth]{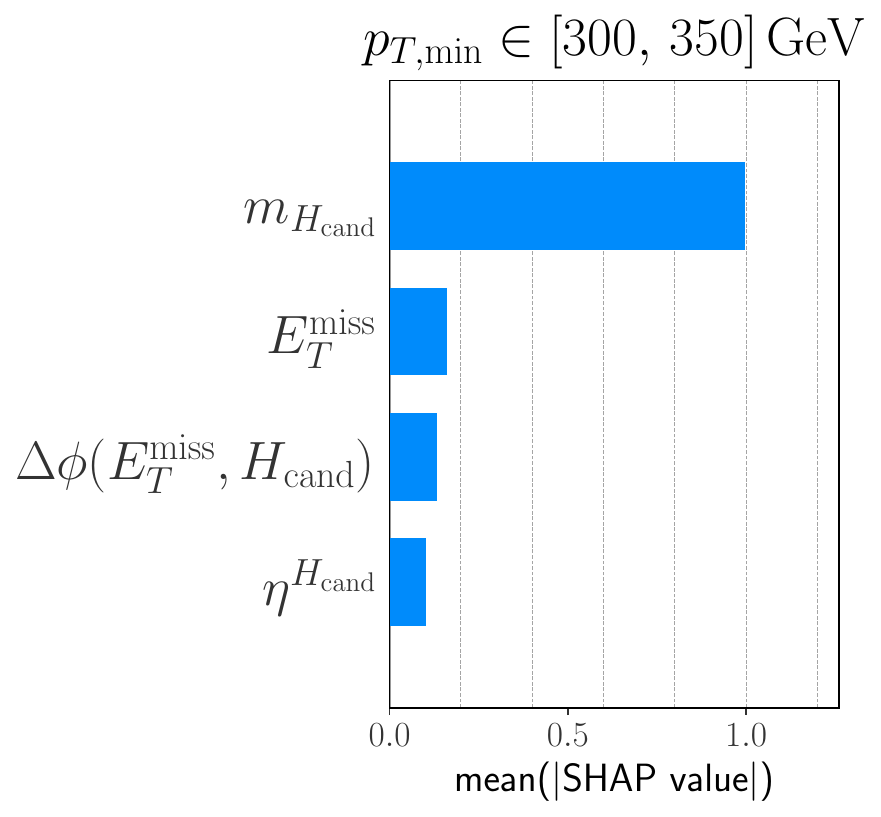} \\
	\vspace{0.3cm}
    \includegraphics[width=0.485\linewidth]{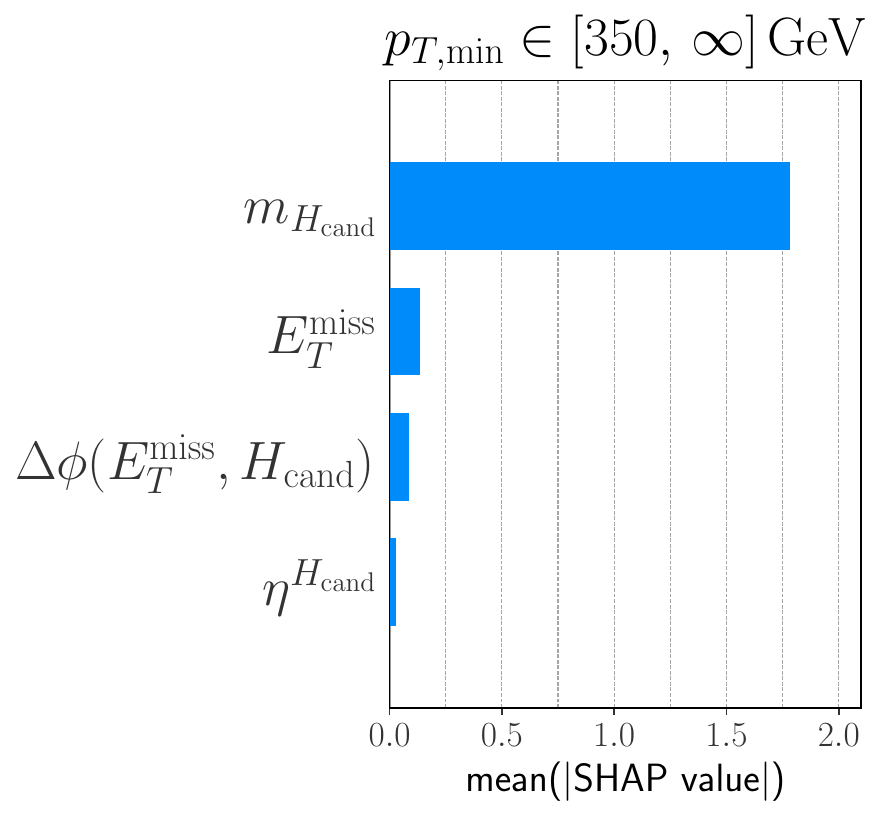}\hfill
 	\caption{Averages of the absolute SHAP values of the different input features used by the BDTs in the boosted 0-lepton category. The top-left panel corresponds to the $[0,\,300]\,\mathrm{GeV}$ bin, the top-right panel to the $[300,\,350]\,\mathrm{GeV}$ bin and the bottom panel to the $[350,\,\infty]\,\mathrm{GeV}$ bin.}
	\label{fig:shap_01} 
\end{figure}

\begin{figure}[tb!]
	\centering
	\includegraphics[width=0.485\linewidth]{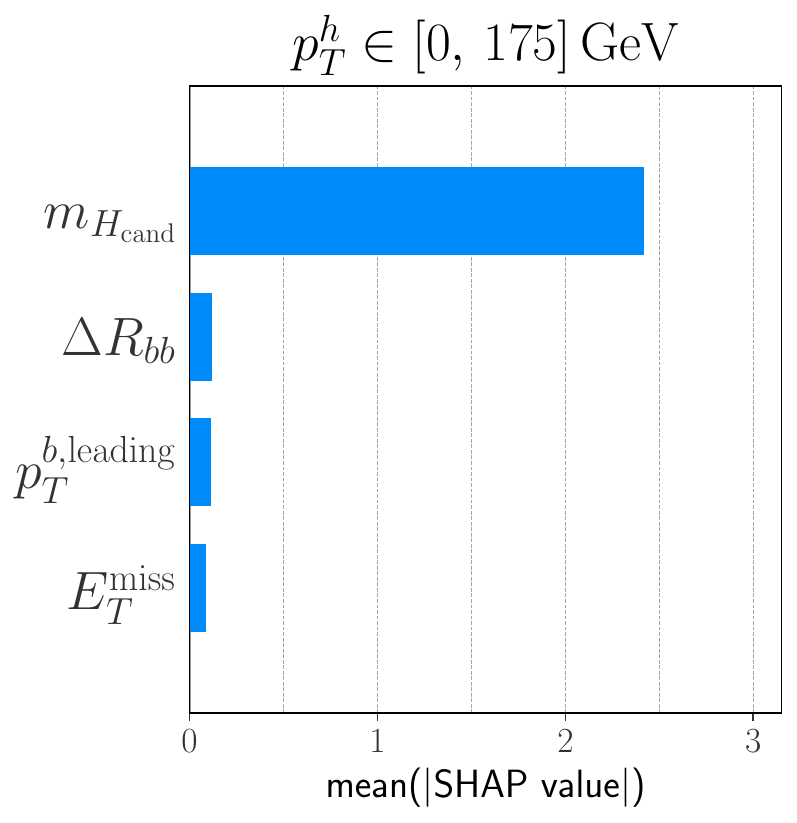} \hfill
	\includegraphics[width=0.485\linewidth]{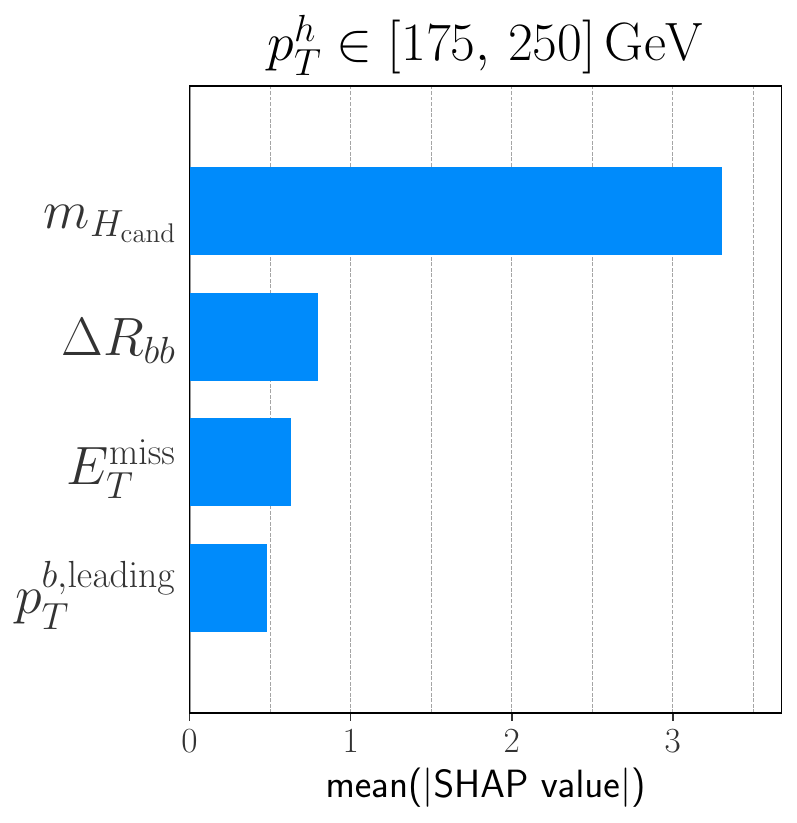} \\
	\vspace{0.3cm}
    \includegraphics[width=0.485\linewidth]{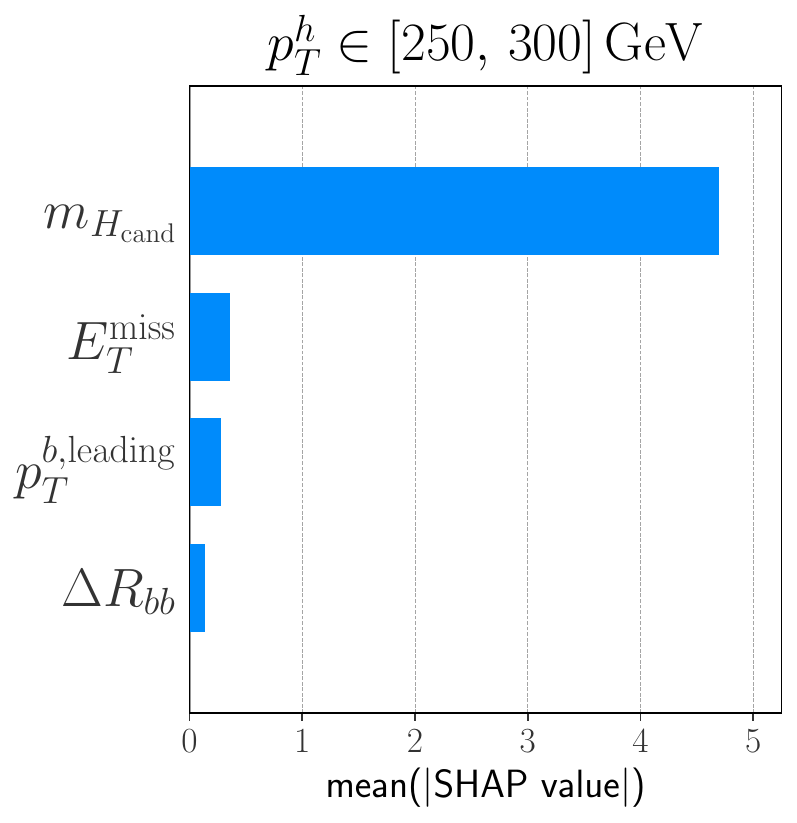}\hfill
 	\caption{Averages of the absolute SHAP values of the different input features used by the BDTs in the resolved 1-lepton category. The top-left panel corresponds to the $[0,\,175]\,\mathrm{GeV}$ bin, the top-right panel to the $[175,\,250]\,\mathrm{GeV}$ bin and the bottom panel to the $[250,\,\infty]\,\mathrm{GeV}$ bin.}
	\label{fig:shap_10} 
\end{figure}

\begin{figure}[tb!]
	\centering
	\includegraphics[width=0.485\linewidth]{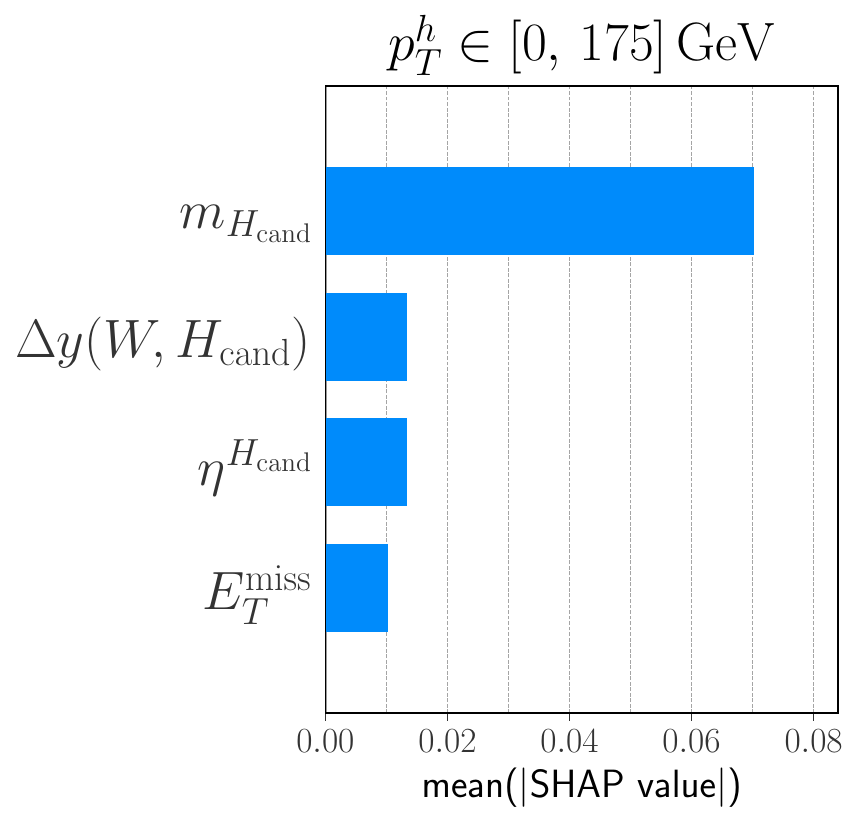} \hfill
	\includegraphics[width=0.485\linewidth]{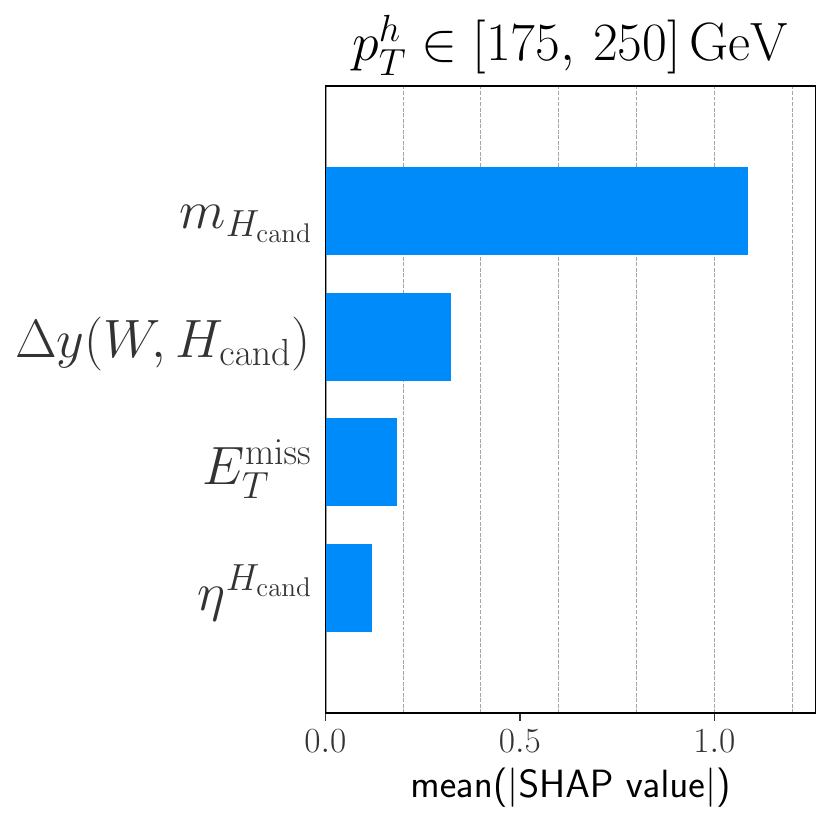} \\
	\vspace{0.3cm}
    \includegraphics[width=0.485\linewidth]{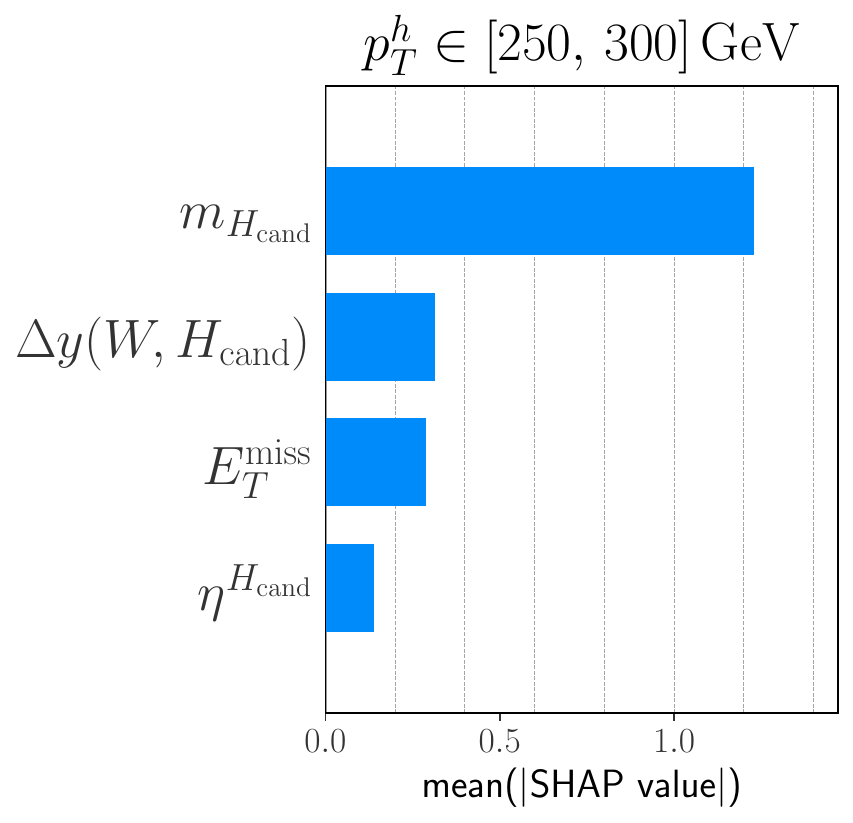}\hfill
	\includegraphics[width=0.485\linewidth]{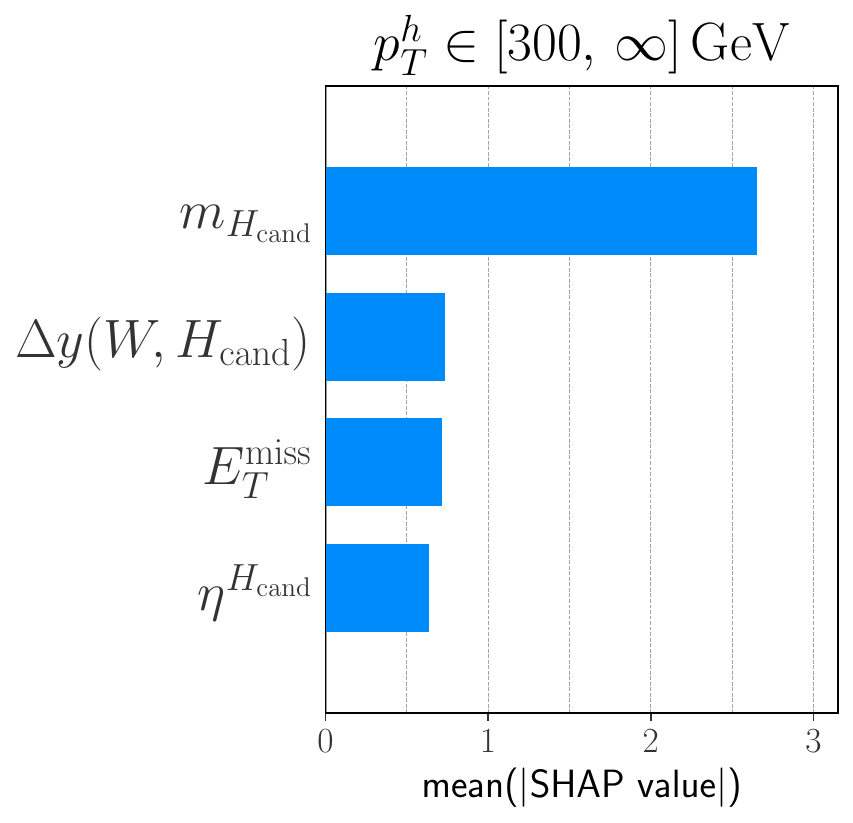}
 	\caption{Averages of the absolute SHAP values of the different input features used by the BDTs in the boosted 1-lepton category. The top-left panel corresponds to the $[0,\,175]\,\mathrm{GeV}$ bin, the top-right panel to the $[175,\,250]\,\mathrm{GeV}$ bin, the bottom-left panel to the $[250,\,300]\,\mathrm{GeV}$ bin and the bottom-right panel to the $[300,\,\infty]\,\mathrm{GeV}$ bin.}
	\label{fig:shap_11} 
\end{figure}

\begin{figure}[tb!]
	\centering
	\includegraphics[width=0.485\linewidth]{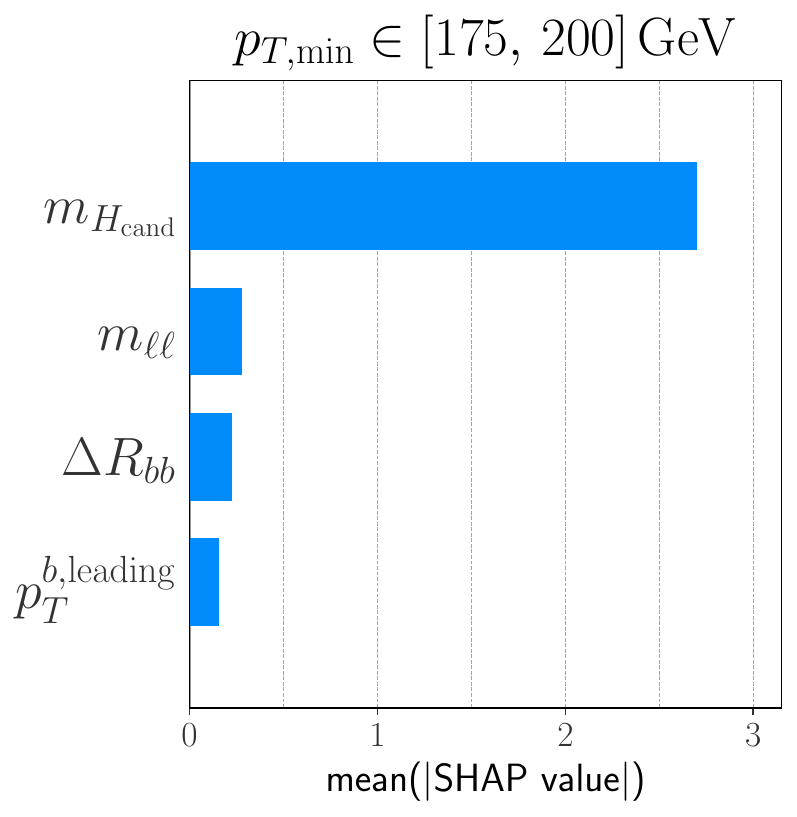} \hfill
	\includegraphics[width=0.485\linewidth]{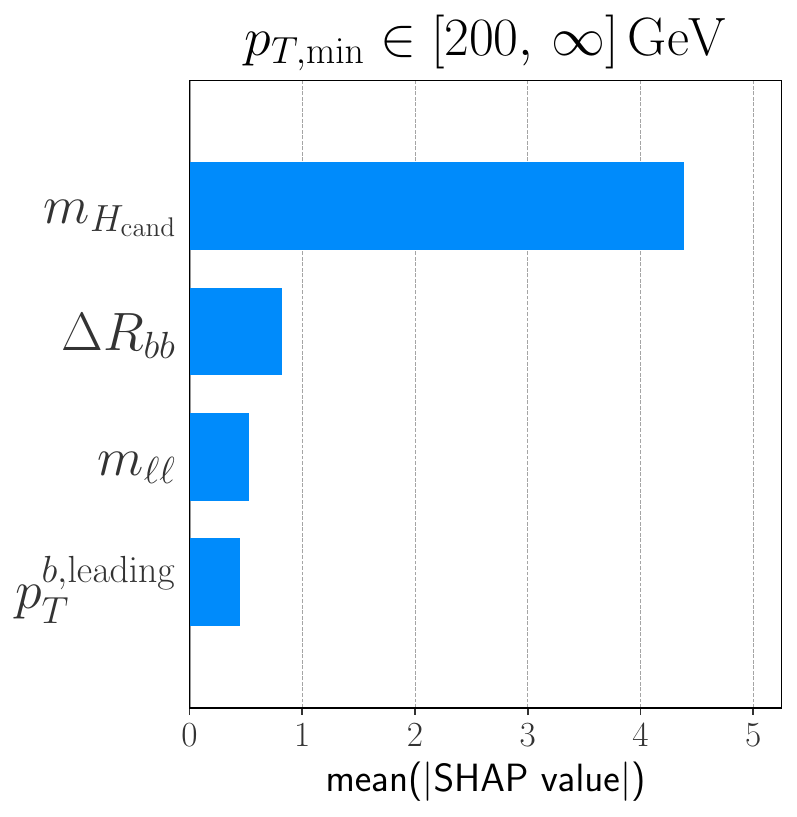} \\
 	\caption{Averages of the absolute SHAP values of the different input features used by the BDTs in the resolved 2-lepton category. The left panel corresponds to the $[175,\,200]\,\mathrm{GeV}$ bin and the right panel to the $[200,\,\infty]\,\mathrm{GeV}$ bin.}
	\label{fig:shap_20} 
\end{figure}

\begin{figure}[tb!]
	\centering
	\includegraphics[width=0.485\linewidth]{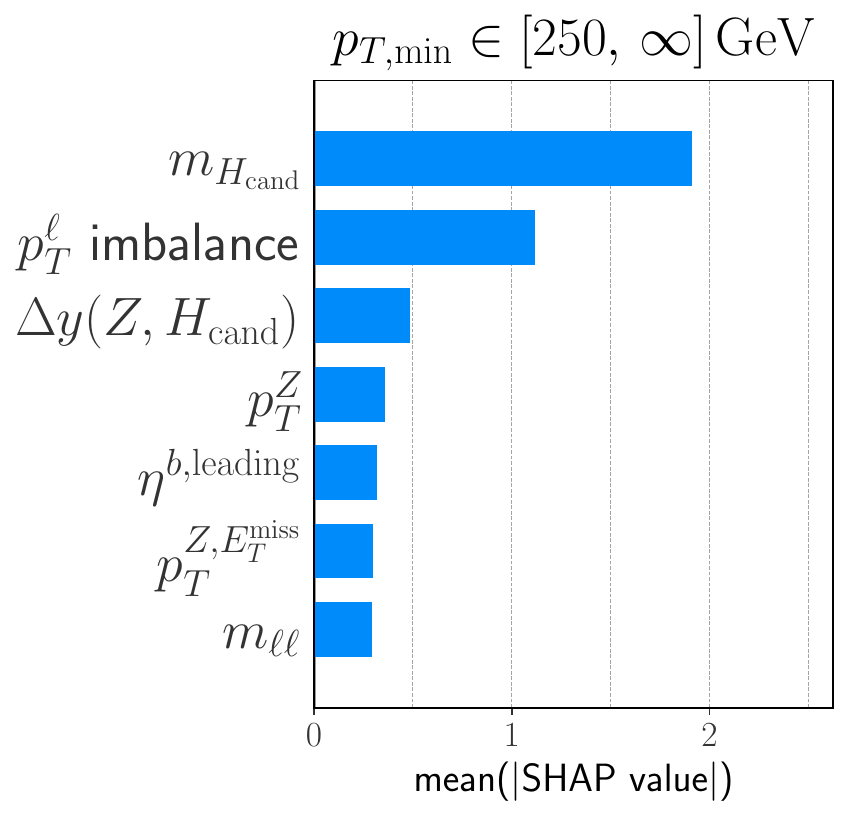}
 	\caption{Averages of the absolute SHAP values of the different input features used by the BDT in the boosted 2-lepton category.}
	\label{fig:shap_21} 
\end{figure}

In this section, we would like to tackle a point of criticism that is commonly direct towards ML-based approaches to particle-physics analyses. Such ML-based analyses are often regarded as black boxes whose internal workings are difficult to understand and therefore hard to learn from. \\
However, we would like to highlight how these algorithms can be demystified by exemplarily studying the impact of our BDT-based analysis on the kinematical distributions of the boosted 0-lepton $Vh(\rightarrow b\bar b)$ category and comparing it with the effects of the corresponding cut-and-count-based approach. Furthermore, we will analyse the mean absolute SHAP values of the different input features used in the BDT-based analysis in order to get some insight into how this particular ML algorithm decides whether an event should be accepted or rejected.\\
In figures~\ref{fig:BDTdistros_010}~-~\ref{fig:BDTdistros_012}, we present the kinematical distributions of the BDTs' input features in the boosted 0-lepton category in the different bins. Each plot shows six distributions. The input features are $\eta^{H_\mathrm{cand}}$, $\Delta\phi(E_T^\mathrm{miss},H_\mathrm{cand})$, $m_H$ and $p_T^Z$. For both signal- (blue) and background events (green), we display the number of unweighted Monte-Carlo events just after the preselection cuts as solid lines, after additionally applying the BDT analysis as dotted lines and after instead applying the rest of the cuts used in the cut-and-count analysis as dash-dotted lines.
\\
\noindent A general observation that can be made for all three bins is that while the cut-and-count analysis typically manifests itself as sharp cuts in the kinematical distributions, the BDT-based analysis does not do this. Instead, we find that the distributions go to zero smoothly, more similarly to how the initial distributions behave. This behaviour is most pronounced in the distributions in the lowest bin displayed in figure~\ref{fig:BDTdistros_010}, especially in the $m_H$ and the $p_T^Z$ distributions. This difference in effects on those distributions can be explained by the fact that a given BDT is modelling a function of the input features that is supposed to represent the probability of a given event to be a signal event. The modelled probability distribution can very well be smooth in the input features and this is what we are observing. The advantage of this strategy over the conventional cut-and-count analysis is that it is potentially able to make better use of the available statistics, since sharp cuts on the distributions typically remove a lot of events from an analysis.\\
We furthermore find that in the last bin, apart from the aforementioned sharp cuts, the distributions sculpted by the BDT analysis are very similar to the ones sculpted by the cut-and-count analysis. In the first two bins however, they differ more. The largest differences can be found in the lowest bin, where the overall number of accepted events is much lower for the cut-and-count analysis, while especially the signal distributions after the BDT analysis follow the initial distributions very closely. This observation is mirrored by the numbers of events expected in the lowest bin at the HL-LHC from the ML-based analysis and the cut-and-count-based analysis reported in tables~\ref{tab:App_sigma_full_Zh_neut_HL_LHC_boos_BDT} and~\ref{tab:App_sigma_full_Zh_neut_HL_LHC_boos}, respectively. The number of expected events for the SM is at least an order of magnitude larger after the BDT-based analysis than after the cut-and-count analysis for both signal and background. In the second bin, the overall number of events is larger for the cut-and-count analysis. The shapes of the distributions sculpted by the two different approaches, however, are rather similar in this bin.\\
We will now discuss figures~\ref{fig:shap_00}~-~\ref{fig:shap_21}, which display the mean absolute values of the SHAP values introduced in section~\ref{sec:ShapleyValues}. We compare these results to the impact of the different cuts in the corresponding cut-and-count analysis given in tables~\ref{tab:cut-flow_NuNu_Boostedhbb}~-~\ref{tab:cut-flow_EllEll_Resolvedhbb}. The comparison shows that the results of the feature-importance analysis for the BDTs mostly agrees with our expectations from the cut-flow tables.
\\
With the exception of the lowest bin in the resolved 0-lepton category, $m_{H_\mathrm{cand}}$ is by far the most important kinematical variable for the discrimination between signal and background. In the lowest bin, the missing transverse momentum surpasses the mass of the Higgs candidate in importance. However, this bin corresponds to the region in phase space, where the cut-and-count analysis has not been optimised and the cut-flow table does not include this low-energy region, since the sensitivity coming from it is very limited. This is the reason why the large discriminatory impact of $E_T^\mathrm{miss}$ seen in the SHAP values for the lowest bin of the resolved 0-lepton category is not reflected by the cut-flow tables. \\
Studying the importance of the other input features, we find that none of them stands out in particular. This is in accord with the cut-flow tables, where we can read off that the impact of most cuts is rather mild. The cut-flow tables show that some cuts, like $\Delta R_{bb}$ in the resolved 2-lepton category, or the $p_T^\ell$ imbalance in the boosted 2-lepton category do exhibit a small discriminatory power, which agrees with their moderate mean absolute SHAP values.\\
This comparison of the cut-flow tables with the mean SHAP values shows that SHAP values are able to provide an understanding about the decision process of a BDT similar to the one a cut-flow table can provide about a cut-and-count analysis. Of course, the SHAP-value analysis we provide here does not give any information about the importance of the different kinematic variables for the rejection or acceptance of events of a specific process like $Zb\bar b$, whereas the cut-flow tables~\ref{tab:cut-flow_NuNu_Boostedhbb}~-~\ref{tab:cut-flow_EllEll_Resolvedhbb} provide this kind of information. However, this lost information can be restored by computing the mean absolute SHAP values for the events of each process individually.

\newpage
\FloatBarrier
\section{Summary and conclusions}\label{sec:conclusionML}
With this study, we showcased how an ML algorithm can be used as an alternative to a conventional cut-and-count analysis by the example of the $V(h\rightarrow b\bar b)$ process. We demonstrated that it is possible to quantify the importance of the different kinematical variables for the predictions using the concept of SHAP values. This technique helps alleviating a common point of criticism directed towards ML applications in the field of particle physics, claiming that it is difficult to interpret and explain their decision processes. \\
To emphasise the point that ML algorithms do not have to be regarded as dubious black boxes, we show that more insight can be gained by, e.g., studying their effects on kinematical distributions. Our findings highlight that, on the one hand, the effects of, e.g., BDTs on the kinematical distributions usually do not manifest as sharp cuts on the distributions as opposed to conventional cut-and-count analyses. This can be advantageous because it can enable a more efficient usage of the available statistics. On the other hand, our study shows that in general, the shapes of the distributions sculpted by the BDTs are not necessarily vastly different from the shapes sculpted by a conventional cut-and-count analysis. This further emphasises the fact that such ML analyses can in fact be made intuitive and interpretable.\\
In terms of quantitative performance, it is difficult to perform a direct comparison between the conventional cut-and-count analysis and the BDT analysis we performed because the BDTs were optimised on a bin-by-bin basis and the cut-and-count analysis was only optimised in detail in the phase-space region of $p_T^V > 200\,$GeV. We find a relative improvement of the bounds of the order $\mathcal{O}(\mathrm{few}\,\%)$ using the BDT-based analysis compared to the cut-and-count analysis, and this advantage could very well be due to the aforementioned difference in methodology.\\
However, we would like to argue that the comparison of tables~\ref{tab:App_sigma_full_Zh_neut_HL_LHC_res_BDT}~-~\ref{tab:App_sigma_full_Zh_lep_HL_LHC_boos_BDT}, containing the number of expected signal- and background events per bin from the ML-based analysis, to the corresponding tables~\ref{tab:App_sigma_full_Zh_neut_HL_LHC_res}~-~\ref{tab:App_sigma_full_Zh_lep_HL_LHC_boos} from the cut-and-count-based analysis indicates that even in the phase-space region of $p_T^V > 200\,$GeV, where the cut-and-count analysis has been carefully optimised, the BDTs are performing better. \\
We highlight this by computing the two quantities $s/b$ and $s/\sqrt{b}$, where $s$ denotes the number of signal events at the HL-LHC assuming a luminosity of $3\,\mathrm{ab}^{-1}$ and $b$ denotes the number of background events. The $\chi^2$ function depends on these two quantities such that larger values of both those quantities yield a larger value of $\chi^2$ and consequently stronger bounds. The quantity $s/b$ is associated with the term that depends on the systematic uncertainty and is therefore more important when the bounds are dominated by systematic uncertainties. The quantity $s/\sqrt{b}$ on the other hand appears in the term that is associated to the statistical uncertainty. We can therefore compute $s/b$ and $s/\sqrt{b}$ bin-by-bin for each category and consider them to be metrics for the performance of the two analysis approaches. We present these quantities for both the cut-and-count analysis and the BDT-based analysis in table~\ref{tab:soverbsoversqrtb}. 
\begin{table}[h!]
\begin{centering}
\setlength{\extrarowheight}{0mm}%
\scalebox{0.95}{
\begin{tabular}{c|c?{1.5pt}c|c||c|c}
\toprule[1.5pt]
\rule[-.5em]{0pt}{.5em}
\multirow{2}{*}{\rule{0pt}{20pt}Category} & $p_T$ bin &\multicolumn{2}{c||}{\rule{0pt}{1.1em}$s/\sqrt{b}$}& \multicolumn{2}{c}{\rule{0pt}{1.1em}$s/b$}\tabularnewline
\cline{3-6} & [GeV] & BDTs& cut-and-count& \rule{0pt}{1.4em} BDTs & cut-and-count \tabularnewline[0.3em]
\midrule[1.5pt]
\multirow{3}{*}{\rule{0pt}{41pt}  $\begin{aligned}&\text{0 - lepton}\\&\text{resolved}\end{aligned}$} & $[0,\,160]$ & \rule{0pt}{1.3em} 25.7 & 3.3 & 0.030 & 0.037 \tabularnewline[0.3em]
\cline{2-6} 
 & $[160,\,200]$ & \rule{0pt}{1.3em} 14.0 &9.5 & 0.26  & 0.13\tabularnewline[0.3em]
\cline{2-6} 
 & $[200,\,250]$ & \rule{0pt}{1.3em} 5.6 & 5.3 & 0.18 & 0.15 \tabularnewline[0.3em]
 \cline{2-6} 
 & $[250,\,\infty]$ & \rule{0pt}{1.3em} 2.2 &2.0  & 0.19 & 0.12 \tabularnewline[0.3em]
\hline
\multirow{3}{*}{\rule{0pt}{30pt} $\begin{aligned}&\text{0 - lepton}\\&\text{boosted}\end{aligned}$} &  $[0,\,300]$ &  \rule{0pt}{1.3em} 9.3 & 5.3 & 0.076 & 0.24\tabularnewline[0.3em]
\cline{2-6} 
 & $[300,\,350]$ & \rule{0pt}{1.3em} 5.1 & 5.3 & 0.26 & 0.24 \tabularnewline[0.3em]
\cline{2-6} 
 &$[350,\,\infty]$ & \rule{0pt}{1.3em} 8.3 & 7.1 & 0.59 & 0.46\tabularnewline[0.3em]
\hline
\multirow{3}{*}{\rule{0pt}{30pt} $\begin{aligned}&\text{1 - lepton}\\&\text{resolved}\end{aligned}$} & $[0,\,175]$ &  \rule{0pt}{1.3em} 32.2 & 13.4 & 0.036 & 0.035\tabularnewline[0.3em]
\cline{2-6} 
 & $[175,\,250]$ & \rule{0pt}{1.3em} 13.9 & 9.6 & 0.21& 0.12\tabularnewline[0.3em]
\cline{2-6} 
 & $[250,\,\infty]$ & \rule{0pt}{1.3em} 3.6 & 2.9 & 0.27 & 0.20 \tabularnewline[0.3em]
\hline
\multirow{3}{*}{\rule{0pt}{41pt}  $\begin{aligned}&\text{1 - lepton}\\&\text{boosted}\end{aligned}$} & $[0,\,175]$ &  \rule{0pt}{1.3em} 1.3 & 0.6 & 0.0054 & 0.016\tabularnewline[0.3em]
\cline{2-6} 
 & $[175,\,250]$ & \rule{0pt}{1.3em} 8.4 & 7.4 & 0.080 & 0.098\tabularnewline[0.3em]
\cline{2-6} 
 & $[250,\,300]$ & \rule{0pt}{1.3em} 7.9 & 6.6 & 0.20& 0.20\tabularnewline[0.3em]
 \cline{2-6} 
 & $[300,\,\infty]$ & \rule{0pt}{1.3em} 4.2 & 3.8 & 0.21 & 0.19\tabularnewline[0.3em]
\hline
\multirow{2}{*}{\rule{0pt}{20pt}  $\begin{aligned}&\text{2 - lepton}\\&\text{resolved}\end{aligned}$} & $[175,\,250]$ &  \rule{0pt}{1.3em} 3.9 & 3.0 & 0.17 & 0.16 \tabularnewline[0.3em]
\cline{2-6} 
 & $[250,\,\infty]$ & \rule{0pt}{1.3em} 2.7 & 2.8 & 0.15 & 0.16\tabularnewline[0.3em]
\hline
\rule{0pt}{21pt}$\begin{aligned}&\text{2 - lepton}\\&\text{boosted}\end{aligned}$ & $[250,\,\infty]$ &  \rule{0pt}{1.3em} 5.1& 5.4& 0.29& 0.28 \tabularnewline[0.3em]
 
\bottomrule[1.5pt]
\end{tabular}
}
\par\end{centering}
\caption{Comparison of $s/b$- and $s/\sqrt{b}$ ratios from the ML analysis with the ones from the cut-and-count analysis in each bin and for the different categories. Here, $s$ denotes the number of signal events at the HL-LHC assuming an integrated luminosity of $3\,\mathrm{ab}^{-1}$ and $b$ denotes the corresponding number of background events.}
\label{tab:soverbsoversqrtb}
\end{table}\\
This shows that the BDTs yield both larger values for $s/\sqrt{b}$ and for $s/b$ than the conventional cut-and-count analysis in the majority of bins for the different categories. Since this better performance by the BDTs is not exclusive to the phase-space region $p_T^V > 200\,$GeV where the cut-and-count analysis has not been carefully optimised, we argue that this can be considered as an indication of the superiority of the BDT approach in terms of quantitative performance in the $Vh(\rightarrow b\bar b)$ analysis overall. Note that the improvement on the bounds is only of the order $\mathcal{O}(\mathrm{few}\,\%)$ despite the large improvements on $s/\sqrt{b}$ and $s/b$ in some of the bins. This can be explained by the fact that the most important bins for the determination of the bounds are the ones exhibiting only a moderate BDT enhancement of $s/\sqrt{b}$ and $s/b$. This is due to the fact that the cut-and-count analysis in ref.~\cite{Bishara:2022vsc0} has been optimised exactly for these most important bins.\\
On this note, we would like to point out that the grid search we performed to optimise the hyperparameters was by no means exhaustive. One could both scan over a more fine-grained grid or a grid that covers a larger hyperparameter space overall, in order to fully exploit the potential performance gain of the ML approach compared to the conventional cut-and-count approach. However, increasing the hyperparameter space is usually accompanied with an increased computation time. In particular for the learning rate, we found it technically difficult to include significantly smaller values into the grid search. This is because smaller learning rates usually require an increased number of training steps, making the grid search even more computationally expensive. A similar effect can potentially be observed for the maximum depth of the trees. However, in our analysis, we found the computation time to be less sensitive to the variations in maximum depth than to the variations in learning rate we applied.\\
Another possibility for a potential performance increase of an ML-based approach to this study of $Vh(\rightarrow b\bar b)$ is the following. In the analysis presented in this paper, we only attempted to improve the efficiency of the discrimination between signal and background after the stage of the boosted-Higgs tagging. In principle, one could however attempt to even replace this step by using lower-level kinematical variables as input for an ML algorithm. However, we would like to point out that due to the increased complexity of the input variables, the hyperparameter-optimisation- and training stage in such an analysis is likely much more involved than the one presented in this chapter. It is conceivable that in order to achieve good sensitivity, one would even have to resort to a different ML algorithm altogether, like e.g. a deep-learning algorithm.\\
Apart from the potential advantage in quantitative performance, another benefit of ML-based approaches is the fact that the work-intensive process of cut-optimisation is replaced by the process of training the ML model. This step is not free of subtleties, e.g. depending on the process to be analysed and depending on the ML algorithm, the hyperparameter optimisation can be more or less intricate. However, we believe that there is much more potential for automatisation compared to the cut-and-count analysis, allowing for human effort to be replaced with computer effort.\\
Even in the study of the $Vh(\rightarrow b\bar b)$ process in particular, there is still more potential for the use of our BDT-based approach. As a proof of concept, we only computed the one-dimensional bounds on the four Wilson coefficients $\chqt$, $\chq$, $\chu$ and $\chd$. However, it could be interesting to also investigate the impact of an ML-based approach on a global analysis turning on several operators at the same time, as we did in the course of the corresponding cut-and-count analysis in ref.~\cite{Bishara:2022vsc0}. We leave this study for future work.

\FloatBarrier
\newpage
\appendix
\section{Signal- and background cross-sections in the ML-analysis}\label{sec:app_tables_ML}
In this appendix, we present the fits for the expected number of signal events and the expected number of background events derived from the BDT-based analysis of $Vh(\rightarrow b\bar b)$ at the HL-LHC assuming an integrated luminosity of $3\,\mathrm{ab}^{-1}$. We give the results for both the resolved and boosted versions of the different $n$-lepton categories.
\vfill
\begin{table}[h!]
	\centering
		\scalebox{.85}{
		\begin{small}
			\begin{tabular}{|c|c|c|c@{\hspace{.25em}}|}
				\hline
				\multicolumn{3}{|c|}{0-lepton channel, resolved, HL-LHC} \tabularnewline
				\hline
				\multirow{2}{*}{
				\hspace{-2.5em}
				\begin{tabular}{c}
				$p_{T,\mathrm{min}}$ bin\\
				$[$GeV$]$
				\end{tabular}
				\hspace{-2.5em}} & \multicolumn{2}{c|}{Number of expected events}\tabularnewline
				\cline{2-3} &  \rule{0pt}{1.15em}Signal & Background \tabularnewline
				\hline
				$[0-160]$ & 
				$\begin{aligned} \phSpa 2.3 &\times 10^4 \, + 6.3\times 10^5 \,\chqt
					+ ( 4 \pm 4)\times 10^3\,\chq + 8.9\times 10^4 \,\chu\\
					 \rule[-1.em]{0pt}{1em}& - (3.6\pm0.3)\times 10^4 \,\chd
					 + 6.1\times 10^6 \,\left(\chqt\right)^{2}
					 + 4.6\times 10^6 \,\left(\chq\right)^{2}\\
					 \rule[-1.em]{0pt}{1em}&
					 +  (2.37 \pm 0.14)\times 10^6\,\left(\chu\right)^{2} +  2.0\times 10^6\,\left(\chd\right)^{2} 
				\end{aligned}$ & $(7.3\pm0.9)\times 10^5$ \tabularnewline \hline
	            $[160-200]$ & $\begin{aligned} \phSpa 730 \,& + (5.34\pm0.03)\times 10^4 \,\chqt
					- ( 1.9 \pm 0.4)\times 10^3\,\chq \\
					\rule[-1.em]{0pt}{1em}&+  (9.8\pm1.3)\times 10^3 \,\chu
					  - (4.6\pm0.6)\times 10^3 \,\chd
					 + 1.2\times 10^6 \,\left(\chqt\right)^{2}\\
					 \rule[-1.em]{0pt}{1em} &+ 9.3\times 10^5 \,\left(\chq\right)^{2}
					 +  ( 5.4\pm 0.4)\times 10^5\,\left(\chu\right)^{2} + 4.0\times 10^5\,\left(\chd\right)^{2} 
				\end{aligned}$ & $2800\pm300$ \tabularnewline
				\hline 
				 $[200-250]$ &  $\begin{aligned} \phSpa 170 \,& + 1.7\times 10^4 \,\chqt
					- ( 9 \pm 3)\times 10^2\,\chq + 3600 \,\chu\\
					 \rule[-1.em]{0pt}{1em}& - (1.35\pm 0.15)\times 10^3 \,\chd
					 + 5.0\times 10^5 \,\left(\chqt\right)^{2}
					 + 4.5\times 10^5 \,\left(\chq\right)^{2}\\
					 \rule[-1.em]{0pt}{1em}&
					 +  2.4 \times 10^5\,\left(\chu\right)^{2} + 1.8\times 10^5\,\left(\chd\right)^{2} 
				\end{aligned}$ & $980\pm160$ \tabularnewline
				\hline 
				 $[250-\infty]$ &  $\begin{aligned} \phSpa 27 \,& + 5.3\times 10^4 \,\chqt
					- ( 3.8 \pm 1.0)\times 10^2\,\chq + (9.4\pm 0.9)\times 10^2 \,\chu\\
					 \rule[-1.em]{0pt}{1em}& - (5.4\pm0.7)\times 10^2 \,\chd
					 + 2.6\times 10^5 \,\left(\chqt\right)^{2}
					 + 2.7\times 10^6 \,\left(\chq\right)^{2}\\
					 \rule[-1.em]{0pt}{1em}&
					 + 1.5 \times 10^5\,\left(\chu\right)^{2} + 9.5\times 10^4\,\left(\chd\right)^{2} 
				\end{aligned}$ & $140\pm70$ \tabularnewline
				\hline 
\end{tabular}
\end{small}}
	\caption{Number of expected signal events as a function of the Wilson coefficients (in units of TeV$^{-2}$) and of total background events in the $Zh \rightarrow \nu \bar{\nu} b\bar{b}$ channel, resolved category, at HL-LHC based on the ML analysis. 
	The Monte-Carlo errors on the fitted coefficients, when not explicitly specified, are $\lesssim \mathrm{few}\,\%$. Note that as opposed to the corresponding tables ~\ref{tab:App_sigma_full_Zh_neut_HL_LHC_res}~-~\ref{tab:App_sigma_full_Zh_lep_HL_LHC_boos}, the functions given for the signal do not include the $\chqt$-$\chq$ mixed terms, since they are just a linear combination of 1D fits.
	}
	\label{tab:App_sigma_full_Zh_neut_HL_LHC_res_BDT}
\end{table}
\vfill
\newpage
\FloatBarrier
\clearpage
\,
\vspace{-1em}
\vfill
\begin{table}[htb!]
	\centering
		\scalebox{.97}{
		\begin{small}
			\begin{tabular}{|c|c|c|c@{\hspace{.25em}}|}
				\hline
				\multicolumn{3}{|c|}{0-lepton channel, boosted, HL-LHC} \tabularnewline
				\hline
				\multirow{2}{*}{
				\hspace{-2.5em}
				\begin{tabular}{c}
				$p_{T,\mathrm{min}}$ bin\\
				$[$GeV$]$
				\end{tabular}
				\hspace{-2.5em}} & \multicolumn{2}{c|}{Number of expected events}\tabularnewline
				\cline{2-3} &  \rule{0pt}{1.15em}Signal & Background \tabularnewline
				\hline
				$[0-300]$ & 
				$\begin{aligned} \phSpa  1130 \,& + 1.4\times 10^5 \,\chqt
					- ( 6.5 \pm 0.8)\times 10^3\,\chq + 2.7\times 10^4 \,\chu\\
					 \rule[-1.em]{0pt}{1em}& - 9900 \,\chd
					 + 5.1\times 10^6 \,\left(\chqt\right)^{2}
					 + 4.4\times 10^6 \,\left(\chq\right)^{2}\\
					 \rule[-1.em]{0pt}{1em}&
					 +   2.5\times 10^6\,\left(\chu\right)^{2} +  1.8\times 10^6\,\left(\chd\right)^{2} 
				\end{aligned}$ & $1.5\times 10^4$ \tabularnewline \hline
	            $[300-350]$ & $\begin{aligned} \phSpa 100 \,& + 2.0\times 10^4 \,\chqt
					- ( 2.1  \pm 0.2 )\times 10^3\,\chq +  4100 \,\chu\\
					 \rule[-1.em]{0pt}{1em}& -1100 \,\chd
					 + 1.2\times 10^6 \,\left(\chqt\right)^{2}
					 +  1.1\times 10^6 \,\left(\chq\right)^{2}\\
					 \rule[-1.em]{0pt}{1em}&
					 + 6.2\times 10^5\,\left(\chu\right)^{2} + 4.3\times 10^5\,\left(\chd\right)^{2} 
				\end{aligned}$ & $400\pm70$ \tabularnewline
				\hline 
				 $[350-\infty]$ &  $\begin{aligned} \phSpa  120\,& + 3.7\times 10^4 \,\chqt
					- ( 3.8 \pm 0.3)\times 10^2\,\chq +8400 \,\chu\\
					 \rule[-1.em]{0pt}{1em}& - 2700 \,\chd
					 + 3.7\times 10^6 \,\left(\chqt\right)^{2}
					 + 3.4\times 10^6 \,\left(\chq\right)^{2}\\
					 \rule[-1.em]{0pt}{1em}&
					 + 2.1 \times 10^6\,\left(\chu\right)^{2} + 1.3\times 10^6\,\left(\chd\right)^{2} 
				\end{aligned}$ & $200\pm 20$ \tabularnewline
				\hline 
\end{tabular}
\end{small}}
	\caption{Number of expected signal events as a function of the Wilson coefficients (in units of TeV$^{-2}$) and of total background events in the $Zh \rightarrow \nu \bar\nu b\bar{b}$ channel, boosted category, at HL-LHC based on the ML analysis. 
	The Monte-Carlo errors on the fitted coefficients, when not explicitly specified, are $\lesssim \mathrm{few}\,\%$. Note that as opposed to the corresponding tables~\ref{tab:App_sigma_full_Zh_neut_HL_LHC_res}~-~\ref{tab:App_sigma_full_Zh_lep_HL_LHC_boos}, the functions given for the signal do not include the $\chqt$-$\chq$ mixed terms, since they are just a linear combination of 1D fits.
	}
	\label{tab:App_sigma_full_Zh_neut_HL_LHC_boos_BDT}
\end{table}
\vfill
\begin{table}[htb!]
	\centering
		\scalebox{.97}{
		\begin{small}
			\begin{tabular}{|c|c|c|c@{\hspace{.25em}}|}
				\hline
				\multicolumn{3}{|c|}{1-lepton channel, resolved, HL-LHC} \tabularnewline
				\hline
				\multirow{2}{*}{
				\hspace{-2.5em}
				\begin{tabular}{c}
				$p_{T}^h$ bin\\
				$[$GeV$]$
				\end{tabular}
				\hspace{-2.5em}} & \multicolumn{2}{c|}{Number of expected events}\tabularnewline
				\cline{2-3} &  \rule{0pt}{1.15em}Signal & Background \tabularnewline
				\hline
				$[0-175]$ & 
				$  \rule[-.85em]{0pt}{2.35em} 2.9 \times 10^4 \, + (8.6\pm4.7)\times 10^5 \,\chqt + 8.8\times 10^6 \,\left(\chqt\right)^{2} $ & $(8.0\pm0.7)\times 10^5$ \tabularnewline \hline
	            $[175-250]$ & 	$  \rule[-.85em]{0pt}{2.35em} 780 \, + (7.3\pm0.2) \times 10^4 \,\chqt + 1.8\times 10^6 \,\left(\chqt\right)^{2} $ & $4500\pm300$ \tabularnewline
				\hline 
				 $[250-\infty]$ &  	$  \rule[-.85em]{0pt}{2.35em} 49 \, + 74005 \,\chqt + 3.0\times 10^5 \,\left(\chqt\right)^{2} $ & $180\pm30$ \tabularnewline
				\hline 
\end{tabular}
\end{small}}
	\caption{Number of expected signal events as a function of the Wilson coefficients (in units of TeV$^{-2}$) and of total background events in the $Wh \rightarrow \nu \ell b\bar{b}$ channel, resolved category, at HL-LHC based on the ML analysis. 
	The Monte-Carlo errors on the fitted coefficients, when not explicitly specified, are $\lesssim \mathrm{few}\,\%$. Note that as opposed to the corresponding tables~\ref{tab:App_sigma_full_Zh_neut_HL_LHC_res}~-~\ref{tab:App_sigma_full_Zh_lep_HL_LHC_boos}, the functions given for the signal do not include the $\chqt$-$\chq$ mixed terms, since they are just a linear combination of 1D fits.
	}
	\label{tab:App_sigma_full_Wh_HL_LHC_res_BDT}
\end{table}
\vfill

\begin{table}[htb!]
	\centering
		\scalebox{.92}{
		\begin{small}
			\begin{tabular}{|c|c|c|c@{\hspace{.25em}}|}
				\hline
				\multicolumn{3}{|c|}{1-lepton channel, boosted, HL-LHC} \tabularnewline
				\hline
				\multirow{2}{*}{
				\hspace{-2.5em}
				\begin{tabular}{c}
				$p_{T}^h$ bin\\
				$[$GeV$]$
				\end{tabular}
				\hspace{-2.5em}} & \multicolumn{2}{c|}{Number of expected events}\tabularnewline
				\cline{2-3} &  \rule{0pt}{1.15em}Signal & Background \tabularnewline
				\hline
				$[0-175]$   &   $  \rule[-.85em]{0pt}{2.35em}  247 \, + (2.2\pm0.7)\times 10^4 \,\chqt + 5.5\times 10^5 \,\left(\chqt\right)^{2} $ & $(6.0\pm1.3)\times 10^4$ \tabularnewline \hline
	            $[175-250]$ & 	$  \rule[-.85em]{0pt}{2.35em}  885\, + 1.0 \times 10^5 \,\chqt + 3.0\times 10^6 \,\left(\chqt\right)^{2} $ & $1.1\times 10^4$ \tabularnewline \hline
	            $[250-300]$ & $  \rule[-.85em]{0pt}{2.35em}  305 \, + 4.6\times 10^4 \,\chqt + 1.8\times 10^6 \,\left(\chqt\right)^{2} $ & $1500\pm 100$ \tabularnewline
				\hline 
				 $[300-\infty]$ &  $  \rule[-.85em]{0pt}{2.35em}  82 \, + 1.9\times 10^4 \,\chqt + 1.3\times 10^6 \,\left(\chqt\right)^{2} $ & $390\pm40$ \tabularnewline
				\hline 
\end{tabular}
\end{small}}
	\caption{Number of expected signal events as a function of the Wilson coefficients (in units of TeV$^{-2}$) and of total background events in the $Wh \rightarrow \nu \ell b\bar{b}$ channel, boosted category, at HL-LHC based on the ML analysis. 
	The Monte-Carlo errors on the fitted coefficients, when not explicitly specified, are $\lesssim \mathrm{few}\,\%$. Note that as opposed to the corresponding tables~\ref{tab:App_sigma_full_Zh_neut_HL_LHC_res}~-~\ref{tab:App_sigma_full_Zh_lep_HL_LHC_boos}, the functions given for the signal do not include the $\chqt$-$\chq$ mixed terms, since they are just a linear combination of 1D fits.
	}
	\label{tab:App_sigma_full_Wh_HL_LHC_boos_BDT}
\end{table}

\begin{table}[htb!]
	\centering
		\scalebox{.92}{
		\begin{small}
			\begin{tabular}{|c|c|c|c@{\hspace{.25em}}|}
				\hline
				\multicolumn{3}{|c|}{2-lepton channel, resolved, HL-LHC} \tabularnewline
				\hline
				\multirow{2}{*}{
				\hspace{-2.5em}
				\begin{tabular}{c}
				$p_{T,\mathrm{min}}$ bin\\
				$[$GeV$]$
				\end{tabular}
				\hspace{-2.5em}} & \multicolumn{2}{c|}{Number of expected events}\tabularnewline
				\cline{2-3} &  \rule{0pt}{1.15em}Signal & Background \tabularnewline
				\hline
				$[0-160]$ & 
				$\begin{aligned} \phSpa 86  \,& + 6500\,\chqt
					+(260 \pm 50) \,\chq + (1400\pm200) \,\chu\\
					 \rule[-1.em]{0pt}{1em}& - (530\pm60) \,\chd
					 + 1.5\times 10^5 \,\left(\chqt\right)^{2}
					 + 1.4\times 10^5 \,\left(\chq\right)^{2}\\
					 \rule[-1.em]{0pt}{1em}&
					 +  (8.3 \pm 0.9)\times 10^4\,\left(\chu\right)^{2} + 7.3 \times 10^4\,\left(\chd\right)^{2} 
				\end{aligned}$ & $510\pm50$ \tabularnewline 
				\hline
				 $[250-\infty]$ &  
				 $\begin{aligned} \phSpa  49 \,& + 5100 \,\chqt
					- (  34\pm 16 )\,\chq + (1100\pm100) \,\chu\\
					 \rule[-1.em]{0pt}{1em}& - (400\pm90) \,\chd
					 + 1.6\times 10^3 \,\left(\chqt\right)^{2}
					 + (1.3\pm 0.2)\times 10^5 \,\left(\chq\right)^{2}\\
					 \rule[-1.em]{0pt}{1em}&
					 +  ( 9.0\pm 0.6)\times 10^4\,\left(\chu\right)^{2} +  6.9\times 10^4\,\left(\chd\right)^{2} 
				\end{aligned}$ & $330\pm40$ \tabularnewline
				\hline 
\end{tabular}
\end{small}}
	\caption{Number of expected signal events as a function of the Wilson coefficients (in units of TeV$^{-2}$) and of total background events in the $Zh \rightarrow \ell^+ \ell^- b\bar{b}$ channel, resolved category, at HL-LHC based on the ML analysis. 
	The Monte-Carlo errors on the fitted coefficients, when not explicitly specified, are $\lesssim \mathrm{few}\,\%$. Note that as opposed to the corresponding tables~\ref{tab:App_sigma_full_Zh_neut_HL_LHC_res}~-~\ref{tab:App_sigma_full_Zh_lep_HL_LHC_boos}, the functions given for the signal do not include the $\chqt$-$\chq$ mixed terms, since they are just a linear combination of 1D fits.
	}
	\label{tab:App_sigma_full_Zh_lep_HL_LHC_res_BDT}
\end{table}

\begin{table}[htb!]
	\centering
		\scalebox{.92}{
		\begin{small}
			\begin{tabular}{|c|c|c|c@{\hspace{.25em}}|}
				\hline
				\multicolumn{3}{|c|}{2-lepton channel, boosted, HL-LHC} \tabularnewline
				\hline
				\multirow{2}{*}{
				\hspace{-2.5em}
				\begin{tabular}{c}
				$p_{T,\mathrm{min}}$ bin\\
				$[$GeV$]$
				\end{tabular}
				\hspace{-2.5em}} & \multicolumn{2}{c|}{Number of expected events}\tabularnewline
				\cline{2-3} &  \rule{0pt}{1.15em}Signal & Background \tabularnewline
				\hline
				$[250-\infty]$ & 
				$\begin{aligned} \phSpa  88 \,& + 1.5\times 10^4 \,\chqt
					- ( 1.4 \pm 0.3 )\times 10^3\,\chq + 3700 \,\chu\\
					 \rule[-1.em]{0pt}{1em}& - (1.21 \pm 0.12)\times 10^3 \,\chd
					 + 8.0\times 10^5 \,\left(\chqt\right)^{2}
					 + 9.2\times 10^5 \,\left(\chq\right)^{2}\\
					 \rule[-1.em]{0pt}{1em}&
					 +   4.9\times 10^5\,\left(\chu\right)^{2} +  3.3 \times 10^6\,\left(\chd\right)^{2} 
				\end{aligned}$ & $300\pm30$ \tabularnewline 
				\hline
\end{tabular}
\end{small}}
	\caption{Number of expected signal events as a function of the Wilson coefficients (in units of TeV$^{-2}$) and of total background events in the $Zh \rightarrow \ell^+ \ell^- b\bar{b}$ channel, boosted category, at HL-LHC based on the ML analysis. 
	The Monte-Carlo errors on the fitted coefficients, when not explicitly specified, are $\lesssim \mathrm{few}\,\%$. Note that as opposed to the corresponding tables~\ref{tab:App_sigma_full_Zh_neut_HL_LHC_res}~-~\ref{tab:App_sigma_full_Zh_lep_HL_LHC_boos}, the functions given for the signal do not include the $\chqt$-$\chq$ mixed terms, since they are just a linear combination of 1D fits.
	}
	\label{tab:App_sigma_full_Zh_lep_HL_LHC_boos_BDT}
\end{table}
\FloatBarrier
\section{Tables for reference from our companion study}\label{sec:app_tables_cutandcount}
In this appendix, we list a collection of tables from our companion paper, ref.~\cite{Bishara:2022vsc0}, for reference. Tables~\ref{tab:cut-flow_NuNu_Boostedhbb}-\ref{tab:cut-flow_EllEll_Resolvedhbb} display the results of the cut-flow analyses for the different categories. In tables~\ref{tab:App_sigma_full_Zh_neut_HL_LHC_res}-\ref{tab:App_sigma_full_Zh_lep_HL_LHC_boos}, we show the expected numbers of signal- and background events at the HL-LHC based on the cut-and-count analysis.
\begin{table}[h!]
\centering
\renewcommand{\arraystretch}{1.15}
\begin{tabular}{c|c|c|c|c|c}
\toprule
Cuts / Eff. & $Zh$  &$Wh$  & $Wb\bar b$  & $Zb\bar b$  & $t\bar t$ \\
\midrule
$0$ $\ell^{\pm}$                                    & $1$    & $0.32$ & $0.34$  & $0.78$      & $0.98$    \tabularnewline
0 UT jets                                           & $0.37$  & $0.036$ & $0.02$    & $0.12$  & $0.011$   \tabularnewline
1 MDT DBT jet                                       & $0.29$ &  $0.026$ & $0.014$  &  $0.048$ &  $0.0018$  \tabularnewline
$\eta_{\max}^{H_\mathrm{cand}}$                     & $0.26$ &  $0.022$ &  $0.012$  &  $0.044$ &  $0.0016$  \tabularnewline
$\Delta \phi (E_T^\mathrm{miss},H_\mathrm{cand})$   & $0.26$ &  $0.022$ &  $0.012$  &  $0.044$ &  $0.0016$  \tabularnewline
$E_{T}^\mathrm{miss}$                               & $0.12$ & $0.007$ & $0.003$  &  $0.013$ &  $0.0005$ \tabularnewline
$m_{H_\mathrm{cand}}$                               & $0.12$ &  $0.007$ &  $0.0008$ &  $0.003$  & $4\cdot10^{-5}$  \tabularnewline
\bottomrule         
\end{tabular}
\caption{Cut-flow for the boosted events in the 0-lepton category at HL-LHC taken from ref.~\cite{Bishara:2022vsc0}. The acceptance regions for charged leptons and jets at the different colliders are defined in the text therein. UT, MDT and DBT stand for untagged, mass-drop-tagged and doubly-b-tagged respectively.}
\label{tab:cut-flow_NuNu_Boostedhbb}
\end{table}

\begin{table}[b!]
\centering
\renewcommand{\arraystretch}{1.15}
\begin{tabular}{c|c|c|c|c|c}
\toprule

Cuts / Eff.& $Zh$  & $Wh$  & $Wb\bar b$  & $Zb\bar b$  & $t\bar t$  \\
\midrule
$0$ $\ell^{\pm}$ & $1$       & $0.32$   & $0.34$ &  $0.78$  & $0.98 $  \tabularnewline
0 UT jets        & $0.37$  & $0.036$  & $0.020$  & $0.12 $  & $0.011 $  \tabularnewline
2 res. $b$-jets  &  $0.028$    &  $0.0027$    &  $0.0016$    &   $0.015 $    &  $ 6\cdot10^{-5}$    \tabularnewline
$\Delta R_{bb}$ & $0.027$  & $0.0024$  & $0.0006$  & $0.0035 $  & $ 1\cdot 10^{-5}$  \tabularnewline
$H_T$ & $0.027$  & $0.0024$  & $0.0006$  & $0.0035 $  & $ 1\cdot 10^{-5}$  \tabularnewline
$p_{T,\min}^{b,\mathrm{leading}}$ & $0.027$  & $0.0024$ & $0.0006$  & $0.0035 $  & $1\cdot10^{-5} $  \tabularnewline
$\Delta \phi (E_T^\mathrm{miss}, H_\mathrm{cand})$ & $0.027$  & $0.0024$ & $0.0006$  & $0.0035 $  & $1\cdot 10^{-5} $  \tabularnewline
$\Delta \phi (b_1, b_2)$ & $0.027$  & $0.0024$  & $0.0006$  & $0.0035 $  & $1\cdot 10^{-5}$  \tabularnewline
$\Delta \phi (E_T^\mathrm{miss}, b\mathrm{-jets})$ & $0.027$  & $0.0024$  & $0.0006$  & $0.0035 $  & $1\cdot 10^{-5}$ \tabularnewline
$E_{T}^\mathrm{miss}$ & $0.027$ &  $0.0024$  & $0.0006$ & $0.0035 $  & $1\cdot 10^{-5}$  \tabularnewline
$m_{H_\mathrm{cand}}$ & $0.027$  & $0.0024$  & $ 3\cdot 10^{-5}$ &  $ 10^{-4}$ & $<10^{-5}$  \tabularnewline
\bottomrule         
\end{tabular}
\caption{Cut-flow for the resolved events in the 0-lepton category at the HL-LHC taken from ref.~\cite{Bishara:2022vsc0}. 
UT stands for untagged.
}
\label{tab:cut-flow_NuNu_Resolvedhbb}
\end{table}
\begin{table}[h]
\begin{centering}
\renewcommand{\arraystretch}{1.1}
\begin{tabular}{c|c|c|c}
\toprule
Cuts / Eff.& $Wh$ &  $Wb\bar b$  & $t\bar t$  \\
\midrule
$1$ $\ell^{\pm}$                                    & $0.66$  & $0.59$   & $0.88$    \tabularnewline
0 UT jets                                           & $0.25$  & $0.075$  & $0.021$   \tabularnewline
1 MDT DBT jet                                       & $0.18$  & $0.051$  & $0.010$   \tabularnewline
$E_{T}^\mathrm{miss}$                               & $0.16$  & $0.043$  & $0.0097$  \tabularnewline
$\eta_{\max}^{H_\mathrm{cand}}$                     & $0.14$  & $0.038$  & $0.0089$  \tabularnewline
$\Delta y (W, H_\mathrm{cand})_\mathrm{max}$        & $0.13$  & $0.030$  & $0.0072$  \tabularnewline
$m_{H_\mathrm{cand}}$                               & $0.13$  & $0.007$  & $0.0005$  \tabularnewline
\bottomrule         
\end{tabular}
\par\end{centering}
\caption{Cut-flow for the boosted events in the 1-lepton category at the HL-LHC taken from ref.~\cite{Bishara:2022vsc0}. 
UT, MDT and DBT stand for untagged, mass-drop-tagged and doubly-b-tagged respectively.}
\label{tab:cut-flow_EllNu_Boostedhbb}
\end{table}
\begin{table}[htb!]
\begin{centering}
\renewcommand{\arraystretch}{1.1}
\begin{tabular}{c|c|c|c}
\toprule
Cuts / Eff.& $Wh$  & $Wb\bar b$  & $t\bar t$  \\
\midrule
$1$ $\ell^{\pm}$                                    &           $0.66$           &       $0.59$             &           $0.88$         \tabularnewline
0 UT jets                                           &           $0.25$           &       $0.075$            &           $0.021$        \tabularnewline
2 res. b-jets                                       & $0.025$  &  $0.006$  &  $0.002$   \tabularnewline
$\Delta R_{bb}$                                     &           $0.025$       &       $0.004$             &        $0.0017$          \tabularnewline
$E_{T}^\mathrm{miss}$                               &           $0.024$       &       $0.003$             &        $0.0016$       \tabularnewline
$p_{T,\min}^{b,\mathrm{leading}}$                   &           $0.024$       &       $0.003$             &        $0.0016$       \tabularnewline
$m_{H_\mathrm{cand}}$                               &           $0.024$       &   $7\cdot10^{-5}$         &        $<5\cdot 10^{-6}$             \tabularnewline
\bottomrule         
\end{tabular}

\par\end{centering}
\caption{Cut-flow for the resolved events in the 1-lepton category at the HL-LHC taken from ref.~\cite{Bishara:2022vsc0}. 
UT stands for untagged.}
\label{tab:cut-flow_EllNu_Resolvedhbb}
\end{table}
\begin{table}[htb!]
\begin{centering}
\renewcommand{\arraystretch}{1.1}
\begin{tabular}{@{}c|c|c@{}}
\toprule
Cuts / Eff.& $Zh$ &   $Zb\bar b$ \\
\midrule
$2$ $\ell^{\pm}$                             & $0.48$  & $0.71$   \tabularnewline
1 MDT DBT jet                                & $ 0.21$ & $0.18 $ \tabularnewline
Leptons                                      & $ 0.21$ & $0.18$  \tabularnewline
$\eta_{\max}^{H_\mathrm{cand}}$              & $ 0.19$ & $0.17 $ \tabularnewline
$\Delta y (Z, H_\mathrm{cand})_\mathrm{max}$ & $0.16$  & $0.10 $ \tabularnewline
$m_{\ell\ell}$                               & $0.15$  & $0.10$  \tabularnewline
max. $p_{T}^{\ell}$ imbalance                & $0.14$  & $0.078$ \tabularnewline
$p_{T,\mathrm{min}}^{Z,E_T^\mathrm{miss}}$   & $0.14$  & $0.078$ \tabularnewline
$p_{T,\mathrm{min}}^Z$                       & $0.14$  & $0.074$ \tabularnewline
$m_{H_\mathrm{cand}}$                        & $0.14 $ & $0.020$ \tabularnewline
\bottomrule         
\end{tabular}
\par\end{centering}
\caption{Cut-flow for the boosted events in the 2-lepton category at the HL-LHC taken from ref.~\cite{Bishara:2022vsc0}. 
MDT and DBT stand for mass-drop-tagged and doubly-b-tagged, respectively.}
\label{tab:cut-flow_EllEll_Boostedhbb}
\end{table}

\begin{table}[htb!]
\begin{centering}
\renewcommand{\arraystretch}{1.1}
\begin{tabular}{@{}c|c|c@{}}
\toprule
Cuts / Eff.& $Zh$  &  $Zb\bar b$  \\
\midrule
$2$ $\ell^{\pm}$                  & $0.48$                 &                  $0.71$ \tabularnewline
2 res. b-jets                     & $0.061$   &  $0.057$    \tabularnewline
$\Delta R_{bb}$                   & $0.052 $                & $ 0.014$                                  \tabularnewline
0 UT jets                         & $0.033 $                & $0.0051 $                 \tabularnewline
Leptons                           & $0.033 $                & $0.0051 $                 \tabularnewline
$p_{T,\min}^{b,\mathrm{leading}}$ & $0.033 $                & $0.0051$                  \tabularnewline
$m_{\ell\ell}$                    & $0.031 $                & $0.0047 $                 \tabularnewline
$m_{H_\mathrm{cand}}$             & $ 0.031$                & $0.0002 $               \tabularnewline
\bottomrule         
\end{tabular}
\par\end{centering}
\caption{Cut-flow for the resolved events in the 2-lepton category at the HL-LHC taken from ref.~\cite{Bishara:2022vsc0}. 
UT stands untagged.}
\label{tab:cut-flow_EllEll_Resolvedhbb}
\end{table}

\begin{table}[h!]
	\centering
		\scalebox{.92}{
		\begin{scriptsize}
			\begin{tabular}{|c|c|c|c@{\hspace{.25em}}|}
				\hline
				\multicolumn{3}{|c|}{0-lepton channel, resolved, HL-LHC} \tabularnewline
				\hline
				\multirow{2}{*}{
				\hspace{-2.5em}
				\begin{tabular}{c}
				$p_{T,\mathrm{min}}$ bin\\
				$[$GeV$]$
				\end{tabular}
				\hspace{-2.5em}} & \multicolumn{2}{c|}{Number of expected events}\tabularnewline
				\cline{2-3} &  \rule{0pt}{1.15em}Signal & Background \tabularnewline
				\hline
				$[0-160]$ & 
				$\begin{aligned} \phSpa 300 \,& + 1120 \,\chqt
					- (39 \pm 47)\,\chq + (155 \pm 39 ) \,\chu - (80\pm30) \,\chd
					 + 1250 \,\left(\chqt\right)^{2}\\
					\rule[-1.em]{0pt}{1em}& + 1010 \,\left(\chq\right)^{2}
					 + (550 \pm 75)\,\left(\chu\right)^{2} +  (400\pm47)\,\left(\chd\right)^{2} - (190 \pm 250)\,\chqt\,\chq
				\end{aligned}$ & $8200\pm3700$ \tabularnewline \hline
	            $[160-200]$ & $\begin{aligned} \phSpa 708 \,& + 3230 \,\chqt
					- (160 \pm 60)\,\chq + ( 596 \pm 49 ) \,\chu - ( 263 \pm 39 ) \,\chd + 4460 \,\left(\chqt\right)^{2} \\
					\rule[-1.em]{0pt}{1em}& + 3340 \,\left(\chq\right)^{2} + 1920 \,\left(\chu\right)^{2} +  1510 \,\left(\chd\right)^{2} - (1400 \pm 340)\,\chqt\,\chq
				\end{aligned}$ & $5500\pm1400$ \tabularnewline
				\hline 
				 $[200-250]$ &  $\begin{aligned} \phSpa 195 \,& + 1160 \,\chqt
					- (55\pm22)\,\chq + (223\pm15) \,\chu - (89\pm13) \,\chd + 2075 \,\left(\chqt\right)^{2}\\
					\rule[-1.em]{0pt}{1em}& + 1750 \,\left(\chq\right)^{2} + 955 \,\left(\chu\right)^{2} +  698\,\left(\chd\right)^{2} - (430 \pm 150)\,\chqt\,\chq
				\end{aligned}$ & $1340\pm90$ \tabularnewline
				\hline 
				 $[250-\infty]$ &  $\begin{aligned} \phSpa 33 \,& + 312 \,\chqt
					-  (32\pm10)\,\chq + (66\pm7) \,\chu - (26\pm6) \,\chd + 1020 \,\left(\chqt\right)^{2}\\
					\rule[-1.em]{0pt}{1em}& + 907 \,\left(\chq\right)^{2} + 517\,\left(\chu\right)^{2} +  351\,\left(\chd\right)^{2} - (360 \pm 85)\,\chqt\,\chq
				\end{aligned}$ & $265\pm37$ \tabularnewline
				\hline 
\end{tabular}
\end{scriptsize}}
	\caption{Expected number of signal events as a function of the Wilson coefficients (in units of TeV$^{-2}$) and total background events in the resolved 0-lepton category, at the HL-LHC based on the cut-and-count analysis in ref.~\cite{Bishara:2022vsc0}. 
	The Monte-Carlo errors on the fitted coefficients, when not explicitly specified, are $\lesssim \mathrm{few}\,\%$.
	}
	\label{tab:App_sigma_full_Zh_neut_HL_LHC_res}
\end{table}
\vfill
\begin{table}[htb!]
	\centering
		\scalebox{.93}{
			\begin{scriptsize}
			\begin{tabular}{|c|c|c|c@{\hspace{.25em}}|}
				\hline
				\multicolumn{3}{|c|}{0-lepton channel, boosted, HL-LHC} \tabularnewline
				\hline
				\multirow{2}{*}{
				\hspace{-2.5em}
				\begin{tabular}{c}
				$p_{T,\mathrm{min}}$ bin\\
				$[$GeV$]$
				\end{tabular}
				\hspace{-2.5em}} &
				\multicolumn{2}{c|}{Number of expected events}\tabularnewline
				\cline{2-3} &  \rule{0pt}{1.15em}Signal & Background \tabularnewline
				\hline
				$[0-300]$ & 
				$\begin{aligned} \phSpa
				118 \,& + 1175 \,\chqt
					- (17 \pm 20) \,\chq + 248 \,\chu - (123\pm13) \,\chd + 3479 \,\left(\chqt\right)^{2}\\
					& + 3064 \,\left(\chq\right)^{2}
					\rule[-1.em]{0pt}{1em}
					+ 1724 \,\left(\chu\right)^{2} +  1300 \,\left(\chd\right)^{2} - (1190 \pm 171)\,\chqt\,\chq
				\end{aligned}$ & $492\pm50$\tabularnewline
				\hline 
				 $[300-350]$ &  $\begin{aligned} \phSpa 117 \,& + 1423 \,\chqt
					- (123 \pm 21)\,\chq + 272 \,\chu - (77\pm13) \,\chd + 5222 \,\left(\chqt\right)^{2}\\
					\rule[-1.em]{0pt}{1em}& + 4643 \,\left(\chq\right)^{2} + 2670\,\left(\chu\right)^{2} +  1810 \,\left(\chd\right)^{2} - (1670 \pm 191)\,\chqt\,\chq
				\end{aligned}$ & $492\pm43$ \tabularnewline
				\hline
				 $[350-\infty]$ &  $\begin{aligned} \phSpa 111 \,& + 2115 \,\chqt
					- (217\pm15)\,\chq + 489 \,\chu - (162\pm9) \,\chd + 12820 \,\left(\chqt\right)^{2}\\
					\rule[-1.em]{0pt}{1em}& + 11790 \,\left(\chq\right)^{2} + 7060\,\left(\chu\right)^{2} +  4650\,\left(\chd\right)^{2} - 4700\,\chqt\,\chq
				\end{aligned}$ & $243\pm16$ \tabularnewline
				\hline
\end{tabular}
\end{scriptsize}}
	\caption{Expected number of signal events as a function of the Wilson coefficients (in units of TeV$^{-2}$) and total background events in the boosted 0-lepton category, at the HL-LHC based on the cut-and-count analysis in ref.~\cite{Bishara:2022vsc0}. 
	The Monte-Carlo errors on the fitted coefficients, when not explicitly specified, are $\lesssim \mathrm{few}\,\%$.
	}
	\label{tab:App_sigma_full_Zh_neut_HL_LHC_boos}
\end{table}

\begin{table}[h!]
		\centering
        \begin{small}
			\begin{tabular}{|@{\hspace{.35em}}c|c|c@{\hspace{.5em}}|}
				\hline
				\multicolumn{3}{|c|}{1-lepton channel, resolved, HL-LHC} \tabularnewline
				\hline
				\multirow{2}{*}{$p_{T}^{h}$ bin [GeV]} & \multicolumn{2}{c|}{Number of expected events}\tabularnewline
				\cline{2-3} &  \rule{0pt}{1.15em}Signal & Background \tabularnewline
				\hline
				$[0-175]$ & $\rule[-.85em]{0pt}{2.35em} 5100 \, + 14900 \,\chqt + 12800 \,\left(\chqt\right)^{2}$ & $144000\pm9800$\tabularnewline
				\hline 
				$[175-250]$ & $\rule[-.85em]{0pt}{2.35em} 780 \, + 4400 \,\chqt + 6600 \,\left(\chqt\right)^{2}$ & $6550$\tabularnewline
				\hline 
                $[250-\infty]$ & $\rule[-.85em]{0pt}{2.35em} 41 \, + 380 \,\chqt + 950 \,\left(\chqt\right)^{2}$ & $203\pm35$\tabularnewline
				\hline
		\end{tabular}
\end{small}
	\caption{Expected number of signal events as a function of the Wilson coefficients (in units of TeV$^{-2}$) and total background events in the resolved 1-lepton category, at the HL-LHC based on the cut-and-count analysis in ref.~\cite{Bishara:2022vsc0}. 
	The Monte-Carlo errors on the fitted coefficients, when not explicitly specified, are $\lesssim \mathrm{few}\,\%$.
	}
	\label{tab:App_sigma_full_Wh_HL_LHC_res}
\end{table}
\vfill
\begin{table}[h!]
	\centering
\begin{small}
			\begin{tabular}{|@{\hspace{.35em}}c|c|c@{\hspace{.5em}}|}
				\hline
				\multicolumn{3}{|c|}{1-lepton channel, boosted, HL-LHC} \tabularnewline
				\hline
				\multirow{2}{*}{$p_{T}^{h}$ bin [GeV]} & \multicolumn{2}{c|}{Number of expected events}\tabularnewline
				\cline{2-3} &  \rule{0pt}{1.15em}Signal & Background \tabularnewline
				\hline
				$[0-175]$ & $\rule[-.85em]{0pt}{2.35em} (26\pm6) \, + (154\pm19) \,\chqt + (221\pm20) \,\left(\chqt\right)^{2}$ & $1630$\tabularnewline
				\hline 
				$[175-250]$ & $\rule[-.85em]{0pt}{2.35em} 560 \, + 3770 \,\chqt + 6650 \,\left(\chqt\right)^{2}$ & $5690$\tabularnewline
				\hline 
                $[250-300]$ & $\rule[-.85em]{0pt}{2.35em} 214 \, + 1920 \,\chqt + 4530 \,\left(\chqt\right)^{2}$ & $1046$\tabularnewline
				\hline 
                $[300-\infty]$ & $\rule[-.85em]{0pt}{2.35em} (79\pm5) \, + 1150 \,\chqt + 4700 \,\left(\chqt\right)^{2}$ & $425\pm25$\tabularnewline
				\hline
			\end{tabular}
\end{small}
	\caption{Expected number of signal events as a function of the Wilson coefficients (in units of TeV$^{-2}$) and total background events in the boosted 1-lepton category, at the HL-LHC based on the cut-and-count analysis in ref.~\cite{Bishara:2022vsc0}. 
	The Monte-Carlo errors on the fitted coefficients, when not explicitly specified, are $\lesssim \mathrm{few}\,\%$.
	}
	\label{tab:App_sigma_full_Wh_HL_LHC_boos}
\end{table}

\begin{table}[h]
	\centering
		\begin{small}
			\begin{tabular}{|@{\hspace{.35em}}c|c|c@{\hspace{.5em}}|}
				\hline
				\multicolumn{3}{|c|}{2-lepton channel, resolved, HL-LHC} \tabularnewline
				\hline
				\multirow{2}{*}{$p_{T,\mathrm{min}}$ bin [GeV]} & \multicolumn{2}{c|}{Number of expected events}\tabularnewline
				\cline{2-3} &  \rule{0pt}{1.15em}Signal & Background \tabularnewline
				\hline
				 $[175-200]$ &  $\begin{aligned} \rule{0pt}{1.25em} 57 \,& + 277 \,\chqt
					- (3\pm7)\,\chq + (73\pm5) \,\chu - (19\pm4) \,\chd\\
					& + 402 \,\left(\chqt\right)^{2}+ 403 \,\left(\chq\right)^{2}\\
					\rule[-.65em]{0pt}{1em}& + 238\,\left(\chu\right)^{2} +  172 \,\left(\chd\right)^{2} - (141 \pm 47)\,\chqt\,\chq
				\end{aligned}$ & $361\pm21$ \tabularnewline
				\hline 
				 $[200-\infty]$ &  $\begin{aligned} \rule{0pt}{1.25em} 48 \,& + 299 \,\chqt
					- (5\pm6)\,\chq + (65\pm5) \,\chu - (25 \pm 4) \,\chd\\
					& + 580 \,\left(\chqt\right)^{2}+ 560 \,\left(\chq\right)^{2}\\
					\rule[-.65em]{0pt}{1em}& + 324\,\left(\chu\right)^{2} +  256\,\left(\chd\right)^{2} - ( 110 \pm 49)\,\chqt\,\chq
				\end{aligned}$ & $296 \pm 19$ \tabularnewline
				\hline
		\end{tabular}
		\end{small}
	\caption{ Expected number of signal events as a function of the Wilson coefficients (in units of TeV$^{-2}$) and total background events in the resolved 2-lepton category, at the HL-LHC based on the cut-and-count analysis in ref.~\cite{Bishara:2022vsc0}. 
	The Monte-Carlo errors on the fitted coefficients, when not explicitly specified, are $\lesssim \mathrm{few}\,\%$.}
	\label{tab:App_sigma_full_Zh_lep_HL_LHC_res}
\end{table}
\vfill
\begin{table}[h]
	\begin{center}
		\setlength{\extrarowheight}{0mm}%
		\begin{small}
			\begin{tabular}{|@{\hspace{.35em}}c|c|c@{\hspace{.5em}}|}
				\hline
				\multicolumn{3}{|c|}{2-lepton channel, boosted, HL-LHC} \tabularnewline
				\hline
				\multirow{2}{*}{$p_{T,\mathrm{min}}$ bin [GeV]} & \multicolumn{2}{c|}{Number of expected events}\tabularnewline
				\cline{2-3} &  \rule{0pt}{1.15em}Signal & Background \tabularnewline
				\hline
				 $[250-\infty]$ &  $\begin{aligned} \rule{0pt}{1.25em} 103 \,& + 974 \,\chqt
					-(53\pm11)\,\chq + 231 \,\chu - (79\pm7) \,\chd\\
					& + 2800 \,\left(\chqt\right)^{2}+ 2850 \,\left(\chq\right)^{2}\\
					\rule[-.65em]{0pt}{1em}& + 1660\,\left(\chu\right)^{2} +  1150\,\left(\chd\right)^{2} - (1070 \pm 93)\,\chqt\,\chq
				\end{aligned}$ & $370\pm21$ \tabularnewline
				\hline
		\end{tabular}
		\end{small}
	\end{center}
	\caption{ Expected number of signal events as a function of the Wilson coefficients (in units of TeV$^{-2}$) and total background events in the boosted 2-lepton category, at the HL-LHC based on the cut-and-count analysis in ref.~\cite{Bishara:2022vsc0}. 
	The Monte-Carlo errors on the fitted coefficients, when not explicitly specified, are $\lesssim \mathrm{few}\,\%$.}
	\label{tab:App_sigma_full_Zh_lep_HL_LHC_boos}
\end{table}

\clearpage

\bibliographystyle{JHEP.bst}
\bibliography{references}
\end{document}